\documentstyle[12pt,epsf]{article}
\begin{document}
  \begin{flushright}
    KANAZAWA-01-08\\
    July 2001
  \end{flushright}

\begin{center}
{\Large\bf 
Muon anomalous magnetic dipole moment \\
         in the $\mu$-problem solvable extra U(1) models}
\end{center} 

\vspace{1cm}
\begin{center} {\sc Yasuhiro Daikoku} 
\end{center}

\begin{center}
{\em 
Institute for Theoretical Physics, 
Kanazawa  University, 
Kanazawa 920-1192, Japan
}
\end{center}

\vspace{1cm}
\begin{center}
{\sc\large Abstract}
\end{center}

\noindent
The recent measurement of the muon anomalous magnetic dipole
            moment shows 2.6$\sigma$ deviation of $a_\mu$ from
            the standard model prediction which can be explained
            by a chargino-sneutrino loop correction in the supersymmetric
	    models.  In this paper we
	    consider extra U(1) models where the $\mu$ parameter is
	    radiatively generated. This model predicts the sign of $\mu$
	    is positive in wide parameter regions. But even 
	    the 2$\sigma$ constraint causes a serious contradiction
	    to the experimental bound of the extra neutral gauge boson mass.
	    Although a minimal supergravity scenario is ruled out, a very
	    small window is remained as an allowed region for
	    a no-scale model with non-universal gaugino masses.

\vspace{1cm}
\noindent

\newpage

\def\vol#1#2#3{{\bf {#1}} ({#2}) {#3}}
\def\NP{Nucl.~Phys. }
\def\PL{Phys.~Lett. }
\def\PR{Phys.~Rev. }
\def\PRP{Phys.~Rep. }
\def\PRL{Phys.~Rev.~Lett. }
\def\PTP{Prog.~Theor.~Phys. }
\def\MPL{Mod.~Phys.~Lett. }
\def\IJMP{Int.~J.~Mod.~Phys. }
\def\JETP{Sov.~Phys.~JETP }
\def\JP{J.~Phys. }
\def\th#1{hep-th/{#1}}
\def\ph#1{hep-ph/{#1}}

\def\equ#1{\begin{equation}
#1
\end{equation}
}

      
\section{Introduction}
Supersymmetry is the most simplest solution of the gauge hierarchy problem in
the standard model (SM), which is one of the main subjects of present 
particle physics \cite{nilles1}. The minimal supersymmetric extension 
of the SM (MSSM) predicts a desirable gauge coupling unification 
and may be the remnant of more fundamental theory, i.e.
GUT or string theory.  Another favorable feature of the MSSM
is radiative electroweak symmetry breaking (EWSB). Large top-Yukawa coupling
induces negative Higgs squared mass through the quantum correction
at a low energy scale, which explains naturally why the electroweak symmetry is
broken. Although the MSSM is the most promising extension of the SM, there 
remain some theoretical problems. The famous one is known as the 
$\mu$-problem. The MSSM has a supersymmetric mass term $\mu H_1H_2$ and 
$\mu$ must be $O(M_Z)$ to cause an appropriate scale of EWSB.
However the size of the $\mu$ scale is naively expected 
to be $O(M_{GUT})$ which
is the fundamental energy scale of the MSSM, so we cannot explain why $\mu$ is
so small. It is natural to consider the origin of the $\mu$ scale as some
result of supersymmetry breaking. Otherwise, even if we have understood 
the smallness of the $\mu$ scale, there would remain another problem why 
soft breaking mass parameters and the $\mu$ scale highly degenerate.

Recently, the new measurement of the muon anomalous magnetic dipole
moment (AMDM)
$a^{exp}_\mu=11 659 202(14)\times 10^{-10}$ shows that it is 2.6$\sigma$ 
away from the standard-model prediction $a^{SM}_\mu$ \cite{g-2,g-2paper}. 
It is interesting to consider this deviation as a new physics effect,
especially supersymmetry (SUSY). The required SUSY contribution is
\begin{equation}
a^{SUSY}_\mu=a^{exp}_\mu-a^{SM}_\mu=43(16)\times 10^{-10},
\end{equation}
which has a positive sign. In the case of the MSSM the dominant contribution
to $a^{SUSY}_\mu$ comes from chargino-sneutrino loop diagrams
\cite{sakai, lopez}. In a small chargino mixing approximation, a chargino
contribution is given by \cite{moroi}
\begin{eqnarray}
a^{\chi^\pm\tilde\nu}_\mu&\sim&\frac{3g^2_2 m^2_\mu\mu M_2 \tan\beta}{16\pi^2
         m^2_{\tilde\nu}(M^2_2-\mu^2)}[f(M^2_2/m^2_{\tilde\nu})
         -f(\mu^2/m^2_{\tilde\nu})], \\
  f(x) &=& (1-x)^{-3}(1-\frac43x+\frac13x^2+\frac23\log{x}),
\end{eqnarray}
where $\tan\beta$ is a ratio of 
the vacuum expectation values (VEVs) of $H_1$ and 
$H_2$. Since $f(x)$ is a simply increasing function, 
$a^{\chi^\pm\tilde\nu}_\mu$ is positive for positive $M_2\mu$. 
The Brookhaven E821 experiment implies the sign of $\mu$ must be
same as $M_2$.

It is very interesting to examine the model which can dynamically
generate both the
appropriate size and the sign of $\mu$ in taking account of $a^{SUSY}_\mu$ 
\cite{martin}. One possibility
of such a model is an introduction of a SM gauge singlet field S 
which replaces $\mu H_1H_2$ by a 
Yukawa type coupling $\lambda SH_1H_2$ in the 
MSSM superpotential. 
If the field S develops a VEV of
order 1 TeV as a result of the negative squared mass of S ($m^2_S < 0$),  
a weak scale value of $\mu$
is dynamically generated through $\mu$=$\lambda<S>$. There are two
scenarios to stabilize the VEV of S. One is to add a cubic term 
$\kappa S^3$ to superpotential. This term breaks PQ symmetry
and prohibits the appearance of a problemaic massless axion. 
To forbid the fundamental $\mu$-term, we must introduce discrete symmetry.
This model is well known as the NMSSM. Another one 
is to extend a gauge symmetry
of the SM. If the field S has a nontrivial charge of the extra gauge symmery,
the potential of a scalar component of S is stabilized by a D-term 
which comes from
this extra gauge multiplet. In this case, 
the massless axion does not apper because of
the Higgs mechanism. The most simplest extension of the 
gauge structure is to add
an extra U(1) symmetry. In this paper we consider this extra U(1) model.
The extra U(1) symmetry forbids the appearance of $\mu H_1H_2$ in the
superpotential without causing the domain wall problem unlike the NMSSM. 
In order to introduce the extra U(1) symmetry,
 additional chiral fermions are needed
for anomaly cancellation. Here, we confine our attention to superstring
inspired $E_6$ models and embed the MSSM matter multiplets in
a {\bf 27} representation of $E_6$ \cite{hewett}.  

In this paper we estimate the muon AMDM taking account of 
a constraint of the extra neutral
gauge boson mass in the correct vacuum. 
The correct vacuum is determined 
as the radiatively induced minimum of the effective potential in the suitable
parameter space. In this approach we use the one-loop effective potential
and solve the relevant renormalization group equations (RGEs) numerically.
In sec.2 we give a short introduction of the extra U(1) model and sec.3 
and sec.4 are devoted to numerical analysis of the muon AMDM.

      
\section{Extra U(1) models}
In this section we define a $\mu$-problem solvable extra U(1) model.
The extra matter contents are determined so as to complete a {\bf 27}
representation of $E_6$, which are listed in Table 1. $E_6$ is a rank 6
group and has two extra U(1)s in addition to the SM gauge group
as its subgroup.
It is decomposed as 
\begin{equation}
 SU(3)_c\times SU(2)_L\times U(1)_Y \times U(1)_\psi\times U(1)_\chi
     \subset E_6.
\end{equation}
At a TeV region only one of
two independent linear combinations of $U(1)_\psi$ and $U(1)_\chi$ are
assumed to remain unbroken and be broken only by the VEV of S. 
In this paper only two extra U(1)
models are considered. They are known as the $\eta$ model and the $\xi$ model and
are defined by
\begin{eqnarray}
 Q_i&=&Q_\psi\cos\theta_i+Q_\chi\sin\theta_i, \\
   \tan\theta_\eta&=&-\sqrt{\frac35}, \nonumber \\
   \tan\theta_{\xi^\pm}&=&\frac{1}{\sqrt{15}}. \nonumber
\end{eqnarray}

The matter contents are given by
\begin{eqnarray}
[3(Q,\bar{U},\bar{D},L,\bar{E})+(H_1,H_2)]_{MSSM}+3(g,\bar{g})
    +2(H_1,H_2)+3(S)+3(N), \nonumber 
\end{eqnarray}
which can be derived from three {\bf 27}s of $E_6$ as is shown in Table 1.
This set satisfies the anomaly free conditions. Unfortunately this
matter multiplet spoils the successful gauge coupling unification in the 
 MSSM. To preserve the unification we must add extra chiral multiplets 
to these in the form of vector representation
\begin{eqnarray}
(H^4_a)+(\bar{H}^4_a), \nonumber
\end{eqnarray} 
where a=1 or 2 and $\bar{H}^4_a$ comes from 
$\bar{\bf 27}$ of $E_6$. At least in the sector of $SU(3)_c\times
SU(2)_L\times U(1)_Y$ these matter contents are the same as [MSSM+
3({\bf 5}+$\bar{\bf 5}$)] where {\bf 5} and $\bar{\bf 5}$ are
the representations of the usual SU(5). \\
\begin{table}
\begin{center}
	\begin{tabular}{|l|ccccccccccc|} \hline
       {fields} & $Q$ & $\bar{U}$ & $\bar{E}$ & $\bar{D}$ & $L$ & $\bar{N}$ &
                  $H_1$ & $\bar{g}$ & $H_2$ & $g$ & S \\ \hline
        SM  & (3,2) & ($\bar{3}$,1) & (1,1) & ($\bar{3}$,1) &
              (1,2) & (1,1) &(1,2) & ($\bar{3}$,1) &(1,2) &(3,1)&(1,1)\\
        Y   & $\frac16$ & $-\frac23$ & 1 & $\frac13$ & $-\frac12$ & 0 &
              $-\frac12$ & $\frac13$ & $\frac12$ & $-\frac13$ & 0 \\
        $\sqrt{\frac25}Q_\psi$ & $\frac16$ & $\frac16$ &
                   $\frac16$ & $\frac16$ &
                   $\frac16$ & $\frac16$ &
                 $-\frac13$ & $-\frac13$ &
                 $-\frac13$ & $-\frac13$ &
                 $\frac23$ \\
        $\sqrt{6}Q_\chi$ & $-\frac{1}{2}$ & $-\frac{1}{2}$ &
                   $-\frac{1}{2}$ & $\frac{3}{2}$ &
                   $\frac{3}{2}$  & $-\frac{5}{2}$ &
                    $-1$ & $-1$ &
                   1 & 1 & 0 \\ \hline
        $Q_\eta$ & $-\frac13$ & $-\frac13$ & $-\frac13$ &
                   $\frac16$ & $\frac16$ & $-\frac56$ &
                   $\frac16$ & $\frac16$ & $\frac23$ & $\frac23$ &
                   $-\frac56$ \\
        $\sqrt{6}Q_{\xi^\pm}$ & $\pm\frac{1}{2}$ & $\pm\frac{1}{2}$ &
                   $\pm\frac{1}{2}$ & $\pm 1$ &
                   $\pm 1$ & 0 & $\mp\frac{3}{2}$ &
                   $\mp\frac{3}{2}$ & $\mp 1$ &
                   $\mp 1$ & $\pm\frac{5}{2}$  \\
                \hline
	\end{tabular} 
         \caption{The charge assignment of 
         extra U(1)s which are derived
         from $E_6$. These 
         charges are normalized as 
         $\sum_{i\in {\bf 27}}Q^2_i=5$ \cite{daikoku}.}
\end{center}
\end{table}
The superpotential of the extra U(1) model is difined by 
\begin{eqnarray}
W&=&h_tQH_2\bar{U}+h_bQH_1\bar{D}+h_\tau LH_2\bar{E}+\sum_{i=1}^3
    k_iSg_i\bar{g}_i+\lambda SH_1H_2 + h_\nu LH_2\bar{N} \\
 &+&\lambda_6 gQQ+\lambda_7 \bar{g}\bar{U}\bar{D}\\
 &+&\lambda_8 g\bar{E}\bar{U}+\lambda_9 \bar{g}LQ+\lambda_{10}g\bar{D}\bar{N}
\end{eqnarray}
where $g_i$ and $\bar{g}_i$ stand for the extra color triplet chiral
superfields.
 We neglect the first and the second generation
Yukawa couplings except for the one including $g$ and $\bar{g}$ in Eq.(6).
Since the coexistence of Eq.(7) and Eq.(8) induces rapid proton
decay at an unacceptable rate,
only one of them can exist.  The model with Eq.(7) and the model 
with Eq.(8) correspond to diquark and leptoquark model, respectively. 
Generally these 
couplings $\lambda_i$ are stringently constrained by 
electroweak rare processes \cite{ellis}.

The existence of multi-generation extra fields brings an ambiguity in 
Eq.(6). The coupling $\lambda$ and $k$ 
can have generation indices for extra fields 
such as $S$, $H_1$, $H_2$, $g$ and $\bar{g}$. On this point we make the
following assumption, for simplicity. For $S^i$ and $H_{1,2}^i$, only one
generation of them can have Yukawa type couplings and get the 
VEVs. In this case
another two generations remain massless after symmetry breaking, which
is not phenomenologically acceptable. But even tiny Yukawa couplings
$\lambda_{ij}\sim O(10^{-2})$ can generate enough size of higgsino mass
through $\lambda_{ij} SH^i_1H^j_2$ and such small couplings do not affect
our RGE analysis and are safely neglected. 
On the other hand, the fermion components
of the remaining $S$ which donot couple to the usual Higgs boson $H_{1,2}$
cannot have a tree level mass, which is problematic
\footnote{If we add $\lambda_kS^kH_1H_2$ to
Eq.(6), the fermion components of $S^k$ become massive through the mixing with
the usual higgsino. 
But $\lambda_k$ must be very large ($\sim O(1)$) to generate a
phenomenologically acceptable mass.}. 
Since the extra colored singlets ($g_i$, 
$\bar{g_i}$) have a diagonal coupling to $S$ as $k_iSg_i\bar{g_i}$, their
fermion components  can get mass through this coupling. This cubic
term plays a crucial role in the breaking of the 
extra U(1) symmetry by driving $m^2_S$
to negative \cite{yamagishi,daikoku}. For the extra chiral multiplets
($H_1^4$, $\bar{H}_1^4$), we must introduce a supersymmetric mass term 
such as  $W_\mu=\tilde{\mu} H^4_1\bar H^4_1$.
If the mass scale of $\tilde{\mu}$ is an appropriate
intermediate scale, the problem of discrepancy between 
$M_{GUT}\sim 2\times 10^{16}$ GeV and $M_{string}\sim 4\times 10^{17}$ GeV 
can be solved. However we are not concerned
with this problem and assume simply $\tilde{\mu}$ is at the weak scale.
As this term does not play an essential role in our analysis, it will be
omitted below.

In this model, there are three new contributions to the muon AMDM. The first
one is a $Z'$ exchange and the second one 
is a leptoquark ($g$, $\bar g$) exchange.
The extra neutral gauge boson gives a small contribution \cite{g-2paper}
\begin{equation}
a^{Z'}_\mu\sim O(1)\times 10^{-11}\left(\frac{m^2_Z}{m^2_{Z'}}\right),
\end{equation}
which is difficult
to be observed for the collider constraint of $m_{Z'}$ ($m_{Z'}>600$ GeV).
The leptoquark contribution is argued in \cite{morris, daikoku2}. 
For the large Yukawa couplings $\lambda_{8,9}\sim O(10^{-1})$, 
this gives a sizable contribution to the muon AMDM. In this paper we donot
consider this effect, for simplicity. The third contribution comes from
the extra U(1) gaugino and the fermion partner of S and  we inculde them in
the neutralino contributions (see appendix A). 

\begin{center}
\begin{figure}[htb]       %
\hspace{10mm}
 \parbox{58mm}{
 \epsfxsize=58mm      %
  \leavevmode
\epsfbox{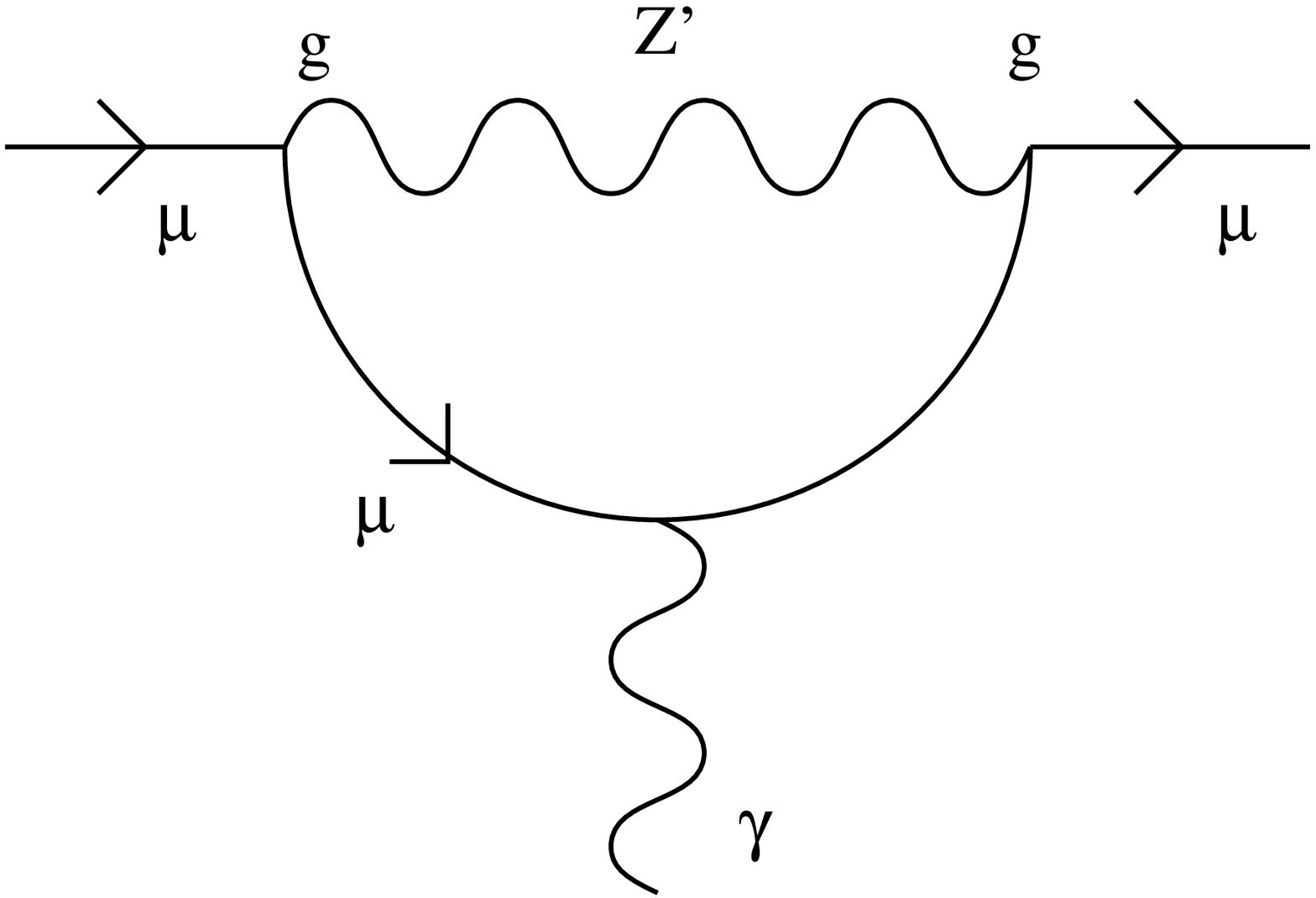}
\vspace{0.5cm}
 }
\hspace{10mm} 
\parbox{58mm}{
 \epsfxsize=58mm      %
 \leavevmode
\epsfbox{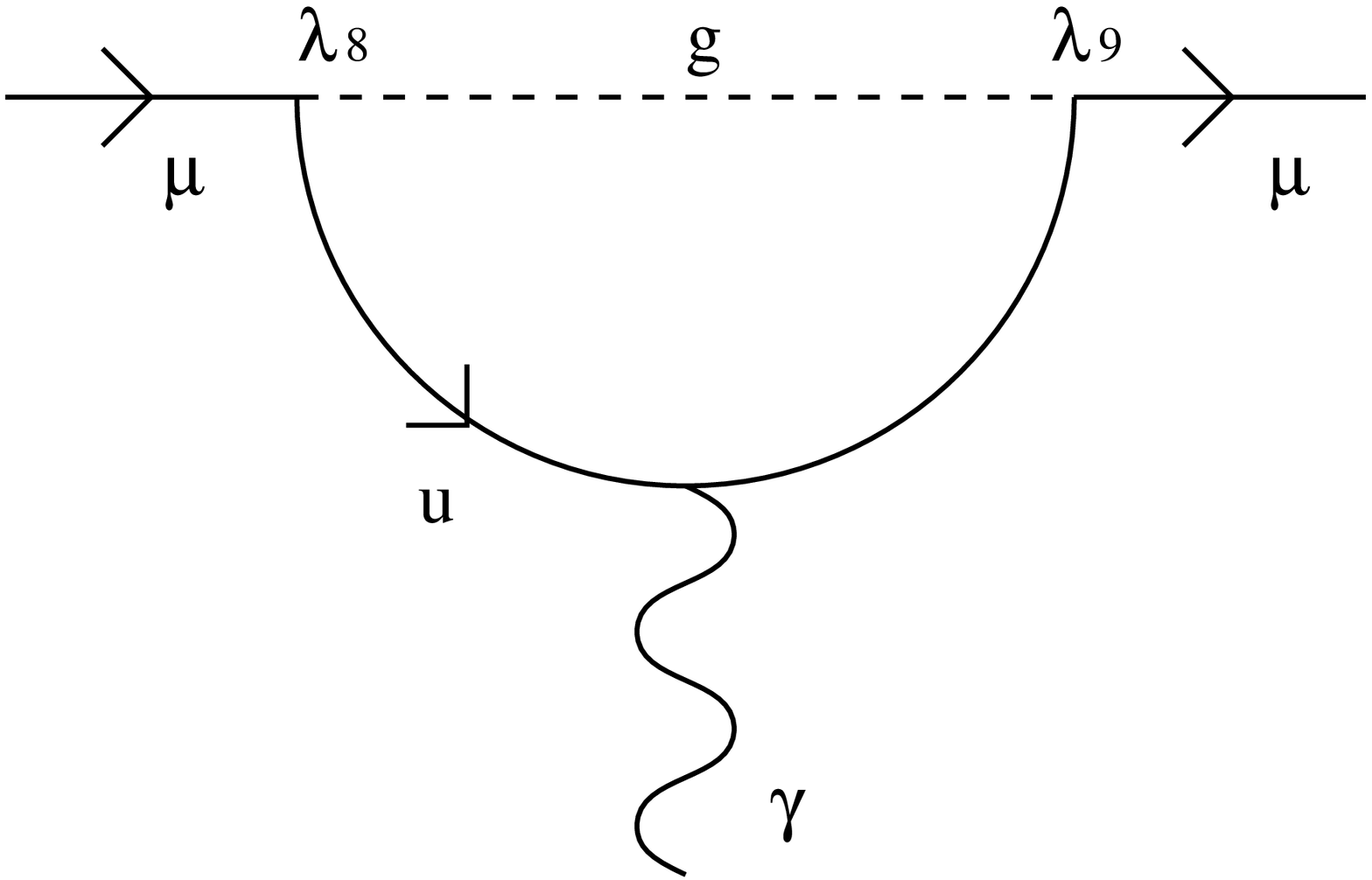}
\vspace{0.5cm}
 }
\caption{The $Z'$ and the leptoquark exchange diagrams 
contribute to the muon AMDM.}
\end{figure}
\end{center}

      
\section{Minimal supergravity case}
Soft supersymmetry breaking parameters are introduced as
\begin{eqnarray}
 {\cal L}_{soft}&=&-\sum_i m^2_{\phi_i}|\phi_i|^2
        -\left(\sum_a \frac12 M_a\lambda_a\lambda_a+h.c.\right) 
        \nonumber \\
        &+&\left(A_\lambda SH_1H_2+A_kS\bar{g}g+A_tQH_2\bar{U}
         +A_bQH_1\bar{D}+A_\tau LH_1\bar{E}+h.c.\right),
\end{eqnarray}
where the first two terms are mass terms of the scalar component $\phi_i$
of each chiral supermultiplet and of gauginos $\lambda_a$. The last
term is a scalar triliner coupling. We use the same notation for the
scalar component as the one of the chiral superfield. In the minimal
supergravity scenario, the values of soft breaking parameters at the
GUT scale are given by
\begin{eqnarray}
m^2_Q=m^2_{\bar U}=\cdots =m^2_S=m^2_0, \nonumber \\
a_t=a_b=\cdots=a_k=A_0, \nonumber \\
M_Y=M_X=M_2=M_3=M_\frac12, \nonumber
\end{eqnarray}
where $A_i=y_ia_i$. It is not so easy to find the phenomenologically
favorable potential minimum
under completely universal soft breaking parameters because the $\mu$ is not a
free parameter unlike the MSSM. To solve the potential minimum condition
exactly, we allow
the non-universality in the region $0.9<m_i/m_0<1.1$ among soft
supersymmetry breaking masses of Higgs scalars. 
Such a small non-universality does not change the mass spectrum 
significantly and  
may give almost the same result as a perfectly universal case.

In our considering models the tree level scalar potential including the
soft supersymmetry breaking terms can be written as
\begin{eqnarray}
V^{(0)}&=&\frac18(g^2_Y+g^2_2)(|H_1|^2-|H_2|^2)^2
         +(|\lambda SH_1|^2+|\lambda SH_2|^2+|\lambda H_1H_2|^2) \nonumber \\
      &+&m^2_1|H_1|^2+m^2_2|H_2|^2+m^2_S|S|^2
       -(a_\lambda\lambda SH_1H_2+h.c.) \nonumber \\
      &+&\frac12 g^2_X(Q_1|H_1|^2+Q_2|H_2|^2+Q_S|S|^2)^2,
\end{eqnarray}
where $Q_1$, $Q_2$, and $Q_S$ are the extra U(1) charges of $H_1$,
$H_2$ and $S$, respectively. The third line is a D-term contribution of 
the extra U(1) and $g_X$ stands for its gauge coupling constant. 
If we replace $\lambda S$ and $a_\lambda$ with
$\mu$ and $B$ and put $g_X=0$, 
this potential becomes the same as the one of the MSSM.  
Since we can take $<H_1>$ and $<H_2>$ positive without loss of generality, 
it is obvious that the global minimum of the scalar potential is 
in a positive $a_\lambda\lambda S$ region. 

The RGEs of the soft breaking parameters are geven in
appendix B. The sign of $a_\lambda$ and $\mu$ are always positive
in natural parameter region at the GUT scale for reproducing a realistic
vacuum, i.e. unbroken $SU(3)_c\times U(1)_{em}$.
 The reason for this is
as follows. In the present extra U(1) model, 
the $\beta$-function of $SU(3)_c$ gauge
coupling is equal to zero, so it always takes a large value
from $M_Z$ to $M_{GUT}$
unlikely the MSSM. The stronger $SU(3)_c$ gauge coupling makes gluino
contribution dominant in RGEs of $a_t$ and $a_k$ and drives them to 
negative. Then the negative $a_t$ and $a_k$ change the sign of
$\beta$-function of $a_\lambda$ negative which forces
$a_\lambda$ take positive before the electroweak scale is reached
\cite{martin}. Finally, the potential minimum condition
forces $\mu$ to take same sign as $a_\lambda$ and the gluino mass.
However, even if we take $A_0$ to be dominant in the RGEs, 
this scenario does not change. As long as
$A_0$ is not much larger, 
the dominant $a_{t,k}$ contributions drive themselves
to zeros, and the gluino dominant condition is again satisfied. To conclude,
the present extra U(1) models predict the positive $\mu$ and 
the positive $a^{SUSY}_\mu$.

Potential minimum condition for Eq.(11) can be written as,
\begin{eqnarray}
m^2_1&=&-\frac14(g^2_Y+g^2_2)(v^2_1-v^2_2)-g^2_XQ_1
      (Q_1v^2_1+Q_2v^2_2+Q_Su^2)-\lambda^2(u^2+v^2_2)+\lambda a_\lambda u
       \frac{v_2}{v_1}, \nonumber \\
m^2_2&=&\frac14(g^2_Y+g^2_2)(v^2_1-v^2_2)-g^2_XQ_2
       (Q_1v^2_1+Q_2v^2_2+Q_Su^2)-\lambda^2(u^2+v^2_1)+\lambda a_\lambda u
       \frac{v_1}{v_2}, \nonumber \\
m^2_S&=&-g^2_XQ_S(Q_1v^2_1+Q_2v^2_2+Q_Su^2)-\lambda^2(v^2_1+v^2_2)
       +\lambda a_\lambda \frac{v_1v_2}{u},
\end{eqnarray}
where $v_1$, $v_2$ and $u$ are the VEVs of $H_1$, $H_2$ and $S$, respectively.
In the extra U(1) models the value of $u$ can be constrained from bellow
by the experimental bounds on the mass of this extra U(1) gauge boson and its
mixing with the ordinary $Z^0$, so that 
we must asuume $u \gg v_1,v_2$. The experimental
constraint of tha extra U(1) gauge boson mass is discussed in
\cite{hagiwara}. In this paper we donot consider the detail structure 
of the $Z-Z'$ mixing, for simplicity.
The third line of Eq.(12) determines the VEV of S such as
\begin{equation}
u\sim\sqrt{-m^2_S/g^2_X Q^2_S},
\end{equation}
and the second line determines the weak scale as
\begin{equation}
-(\lambda^2+Q_2Q_Sg^2_X)u^2-m^2_2\sim\frac12 m^2_Z,
\end{equation}
where the large $\tan\beta$ approximation should be understood. This condition
constrains the allowed range of $\lambda(M_S)$ and $\mu_{eff}$
severely. The first
line of Eq.(12) is written as
\begin{equation}
\lambda u a_\lambda\tan\beta\sim m^2_1+(\lambda^2+g^2_XQ_1Q_S)u^2,
\end{equation}
which can be consistent with
the large $\tan\beta$ solution as far as $m^2_1 \gg m^2_2$ is
satisfied.
This condition makes it difficult to realize the large $\tan\beta$ solution.
From the point of view of the RGE analysis,
the large $\tan\beta$ makes the low energy values of $m^2_1$ and 
$m^2_2$ degenerate due to the same RGE evolution \cite{moroi2}.
However, the degeneracy between $m^2_1$ and $m^2_2$ makes $v_2/v_1$
small at the scalar potential minimum. In this way, the moderate $\tan\beta$
solution is favored for the extra U(1) models (Too large
$\tan\beta$ solution is disfavored).

For more precise estimation, we must take account of the radiative correction
to the potential, as it may make a sizable contribution mainly 
due to the heavy stops.
It is well-known that the one-loop contribution to the effective
potential can be written as
\begin{equation}
V^{(1)}=\frac{1}{64\pi^2}Str {\cal M}^4\left(
    \ln\frac{{\cal M}^2}{\Lambda^2}-\frac32 \right),
\end{equation}
where ${\cal M}^2$ is a matrix of the squared mass of the fields
contributing to the one-loop correction and $\Lambda$ is a renormalization 
point which is taken as $M_S(=1TeV)$. 
In the case of the MSSM, 
this one-loop correction is dominated by top and stop contributions
because of their large Yukawa coupling and the other fields
are irrelevant. In the study of the extra U(1) models $k$ is rather large
and then we should also take account of the effect on ${\cal M}^2$
from the extra colored chiral superfields $g$ and $\bar{g}$.
Mass matrices of these and another sparticles are given in appendix C.

Taking account of experimantal constraints, we get phenomenologically
allowed regions of the parameter space, which are given in the form of
mass bound as \cite{data},
\begin{eqnarray}
&m_{H^\pm}\geq 69 \quad GeV,& \quad m_{\chi^\pm}\geq 72 \quad GeV,
\quad m_{\tilde t^\pm} \geq 86\quad GeV, \nonumber \\
&M_3\geq 180 \quad GeV,& \quad m_{Z'} \geq 600 \quad GeV,
\quad m_{H^0}\geq 114 \quad GeV, \nonumber \\
&m_{\tilde\tau^\pm} \geq 81 \quad GeV,& 
\quad m_{\tilde b^\pm} \geq 75 \quad GeV, \quad m_{g,\tilde g}\geq 220 \quad
GeV,
\end{eqnarray}
where the mass of $Z'$ boson is written by
\begin{equation}
m^2_{Z'}=2g^2_X(Q^2_1v^2_1+Q^2_2v^2_2+Q^2_Su^2)\sim -2m^2_S.
\end{equation}
The explicit formulas of the masses of neutral and charged Higgs bosons
 are given in our previous work \cite{daikoku}. These
mass spectra are mainly governed by $m_0$ and $M_{1/2}$ and
are highly correlated each other. If the mass bounds of $Z'$ 
boson and charginos are
satisfied, the other mass bounds become trivial
\footnote{In all the parameter space, 
the lightest chargino is always wino-like
chargino because of the large $\mu$ parameter.}. Large $\mu_{eff}$
makes a charged Higgs boson heavy and large soft breaking parameters
make sparticles heavy. Both are immediate results of the heavy $Z'$. 
If the chargino mass bound is satisfied, 
we get $M_2=\frac{\alpha_2}{\alpha_3}M_3\sim 0.3M_3>72$ GeV
from the gaugino mass unification relation: 
\begin{equation}
 \frac{M_3}{\alpha_3}=\frac{M_2}{\alpha_2}=k_Y\frac{M_1}{\alpha_Y},
\end{equation}
where $k_Y$ is Kac-Moody level of $U(1)_Y$,   then
the gluino mass bound is trivially satisfied.

For dimensionless parameters
we investigate the parameter region such as 
\begin{equation}
0.4>\lambda(M_S)>0.2, \quad 0.7>k(M_S)>0.4, \quad \tan\beta=(10,20),
\end{equation}
where the value of $\tan\beta$ is given at $M_Z$. At $M_Z$ we convert
the running masses of tau and bottom to
the tau and bottom Yukawa couplings by
\begin{eqnarray}
y_b(M_Z)&=&\hat{m_b}(M_Z)/v_1(M_Z)=2.92GeV/v(M_Z)\cos\beta, \nonumber \\
y_\tau(M_Z)&=&\hat{m_\tau}(M_Z)/v_1(M_Z)=1.74GeV/v(M_Z)\cos\beta, \nonumber
\end{eqnarray}
where $v=175$ GeV.
Using these initial conditions of gauge and Yukawa couplings they are given
at $M_Z$ by \cite{pierce}
\begin{eqnarray}
\alpha_Y(M_Z)&=&0.01698, \nonumber \\
\alpha_2(M_Z)&=&0.03364, \nonumber \\
\alpha_3(M_Z)&=&0.118, \nonumber \\
y_b(M_Z)&=&0.0166/cos\beta, \nonumber \\
y_\tau(M_Z)&=&0.0099/cos\beta. \nonumber
\end{eqnarray}
We run them from $M_Z$ to $M_{top}$ with the RGEs for the SM. 
At $M_{top}$, we define the top-Yukawa coupling by
\begin{equation}
y_t(M_{top})v_2(M_{top})=
      \hat{m}_{pole}(1+\frac{4}{3\pi}\alpha_3(M_{top}))^{-1},
\end{equation}
where $\hat{m}_{pole}=M_{top}=175$ GeV. Finally we run them from $M_{top}$ to
$M_S$ with the RGEs for the two Higgs doublet model \cite{demir},
\begin{eqnarray}
 (2\pi)\frac{\alpha_Y}{dt}&=&7\alpha^2_Y, \\
 (2\pi)\frac{\alpha_2}{dt}&=&-3\alpha^2_2, \\
 (2\pi)\frac{\alpha_3}{dt}&=&-7\alpha^2_3, \\
(2\pi)\frac{Y_t}{dt}&=&Y_t[\frac92Y_t+\frac12Y_b-\frac{17}{12}\alpha_Y
        -\frac{9}{4}\alpha_2-8\alpha_3],\\
 (2\pi)\frac{Y_b}{dt}&=&Y_b[\frac92Y_b+\frac12Y_\tau+\frac12Y_t
         -\frac{5}{12}\alpha_Y
        -\frac{9}{4}\alpha_2-8\alpha_3],\\
 (2\pi)\frac{Y_\tau}{dt}&=&Y_\tau [3Y_b+\frac52 Y_\tau
        -\frac{15}{4}\alpha_Y-\frac94\alpha_2], \\
 (2\pi)\frac{v_1}{dt}&=&\frac{1}{2}v_1[\frac34\alpha_Y+\frac94\alpha_2
         -3Y_b-Y_\tau],\\
(2\pi)\frac{v_2}{dt}&=&\frac{1}{2}v_2[\frac34\alpha_Y+\frac94\alpha_2
         -3Y_t],
\end{eqnarray}
from which we get  initial conditions of dimensionless couplings at
$M_S$. 
For soft supersymmetry breaking parameters we give them at $M_{GUT}$ by
\begin{equation}
 m_0=0.6, \quad 0.18<M_{\frac12}<0.36, \quad -1.2<A_0<1.2 ,
\end{equation}
where these are given by a TeV unit and
the gluino mass lower bound is taken acount of
previously\footnote{In our considering extra U(1) models, the 
$\beta$-function of $M_3$ equals to zero so that $M_3$ is constant.}. 
Out of these parameter regions it is difficult to satisfy the 
potential minimum condition. 
For fixed values of the Yukawa couplings, the parameter set
$(m_0,A_0,M_{\frac12})$ has the only one degree of freedom since they are
imposed by two constraints from Eq.(12). So the SUSY breaking scale is 
represented by only one of them, we take it as $m_0$.
In order to improve the one-loop effective potential 
we use two-loop RGEs for Yukawa and gauge coupling constants and
soft scalar masses and one-loop ones for A-parameters and gaugino masses
from $M_{GUT}$ to supersymmetry breaking scale $M_S$ \cite{yamada}. 
\begin{center}
\begin{figure}[t]       %
\hspace{10mm}
 \parbox{60mm}{
 \epsfxsize=60mm      %
  \leavevmode
\epsfbox{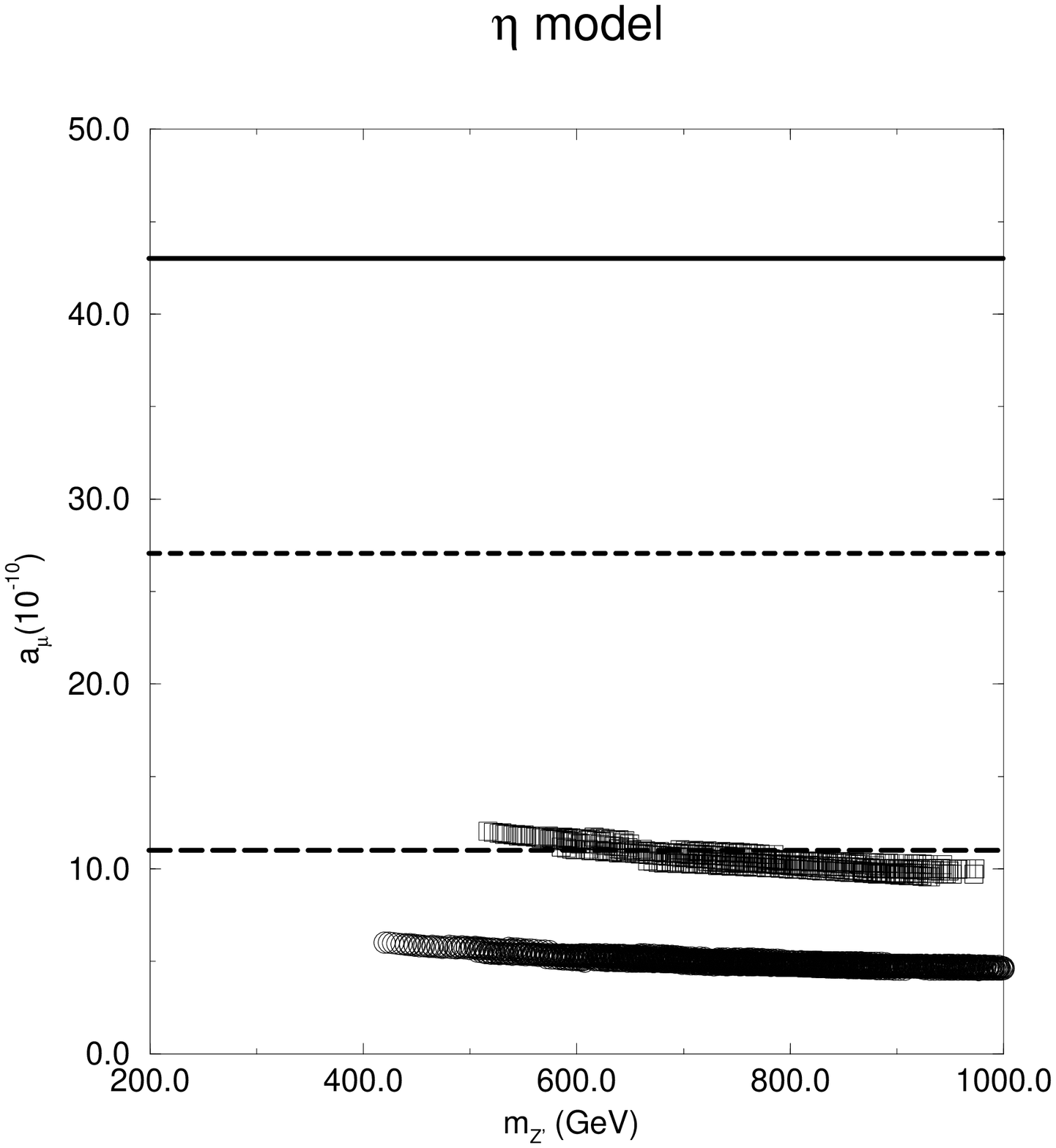}
\vspace{-0.5cm}
 }
\hspace{10mm} 
\parbox{60mm}{
 \epsfxsize=60mm      %
 \leavevmode
\epsfbox{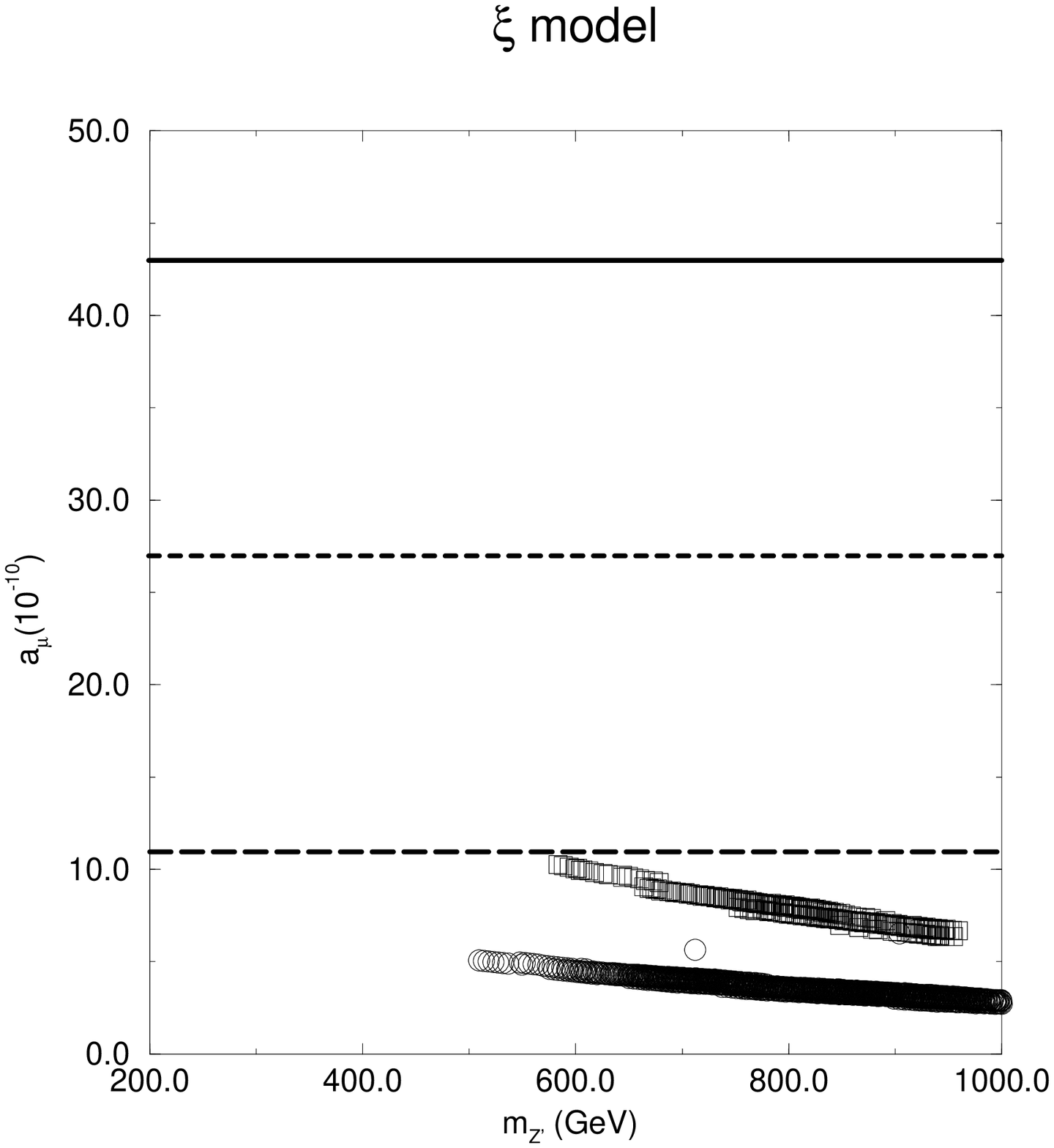}
\vspace{-0.5cm}
 }
\caption{Scatter plots of the radiative symmetry breaking solutions
for the  $\eta$ model and the $\xi^-$ model in the
($a^{SUSY}_\mu$, $m_{Z'}$) plane. 
Solutions for the different values of $\tan\beta$ are classified as
the circles for $\tan\beta=10$ and the squares for $\tan\beta=20$,
respectively. The solid lines stand for
$a^{SUSY}_\mu=43(\times 10^{-10})$ 
and the dashed and the long-dashed lines stand for
$1\sigma$ and $2\sigma$ bounds, respectively.
}
\end{figure}
\end{center}

We evaluate the chargino and neutralino contributions to the muon AMDM
based on the known formula for the MSSM as given in appendix A.
The results are shown in Fig.2. With universal soft breaking terms,
the allowed region never enters inside the $1 \sigma$ bound
$27< a^{SUSY}_\mu<59 \quad (\times 10^{-10})$ either for the $\eta$ model
or the $\xi^-$ model. This does not change 
even in the case where the large $\tan\beta$
enhancement exists.  Although 
there are new contributions from an extra U(1) gaugino and 
a new singlet fermion $\tilde S$, since neutralino 
contributions are always small due to small mixing angle 
of smuon eigenstates, 
they do not play essential role \cite{lopez}.
The main obstacle of inducing the large $a^{SUSY}_\mu$
is due to the chargino mass lower bound 
because the potential minimum condition
favors the small gluino mass. It is shown in Fig.3 
that the chargino mass 
constraint
is sronger than the one of extra neutral gauge boson mass.
In the case of larger $m_0$, sneutrino becomes heavier and 
suppresses $a^{SUSY}_\mu$ more strongly.
On the other hand, in the smaller $m_0$ case, the chargino mass bound
excludes the wider region of the parameter space. In conclusion, the
minimal supergravity scenario is ruled out by the muon AMDM constraint
at the $1 \sigma$ level
if we take account of only the chargino and neutralino loop effects.
The $2 \sigma$ bound $11< a^{SUSY}_\mu<76 \quad (\times 10^{-10})$
gives the upper mass bound of extra neutral gauge boson about 600 GeV
for the $\eta$ model. However, the $\xi$ model is excluded even at the
$2\sigma$ level.

Because of the above argument, it is interesting to consider the
case without the gaugino mass 
universality in order to escape from the chargino mass
constraint. But if we allow the gaugino mass non-universality,
there is no reason why the non-universality of soft scalar masses
and scalar trilinear couplings are forbidden. 
Although such a general non-univeasal case
is interesting, in that case
we must take care of the FCNC constraints and must
invoke some FCNC suppression mechanism, which is out of our
present scope.
\begin{center}
\begin{figure}[htb]       %
\hspace{10mm}
 \parbox{60mm}{
 \epsfxsize=60mm      %
  \leavevmode
\epsfbox{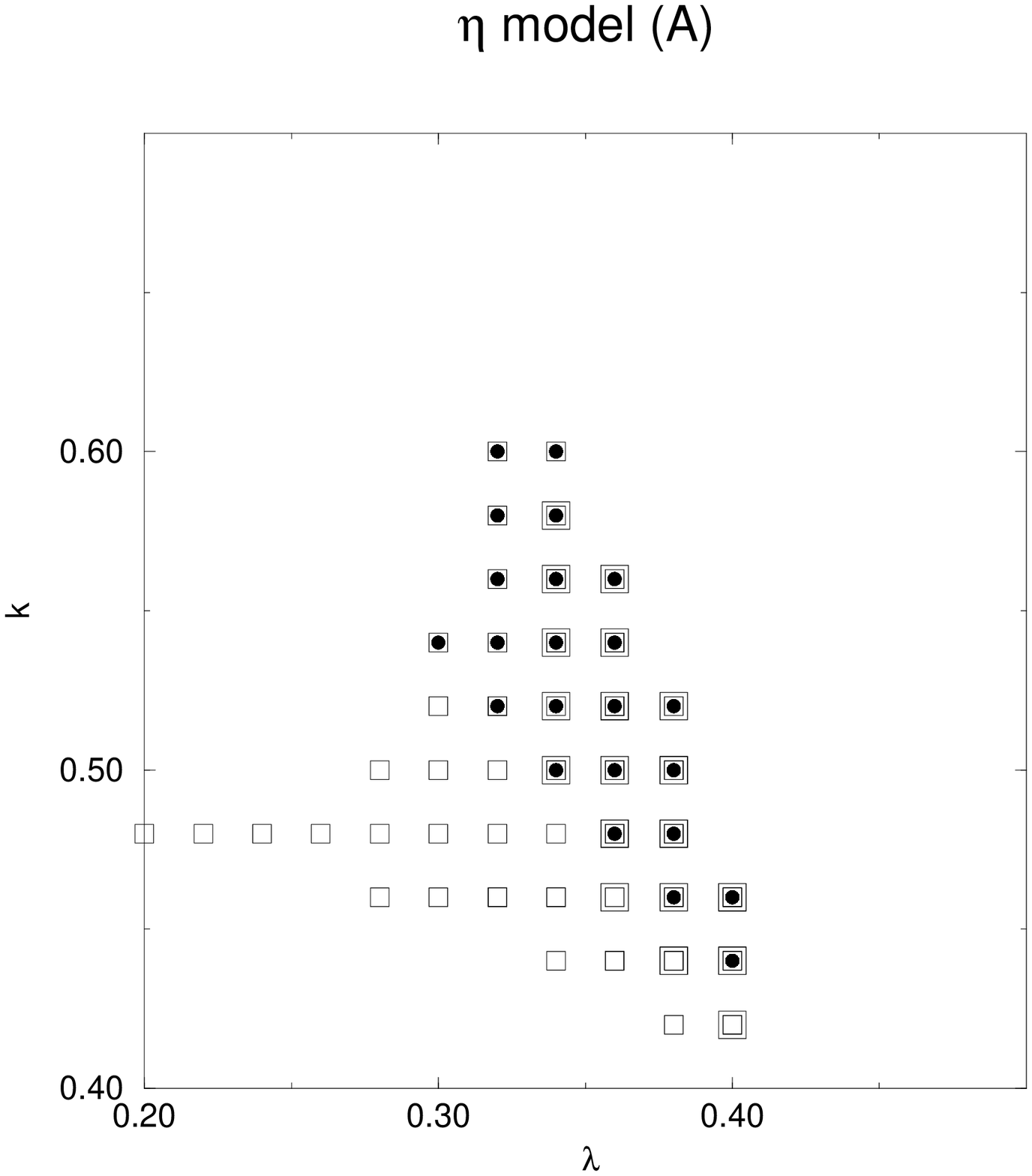}
\vspace{-0.5cm}
 }
\hspace{10mm} 
\parbox{60mm}{
 \epsfxsize=60mm      %
 \leavevmode
\epsfbox{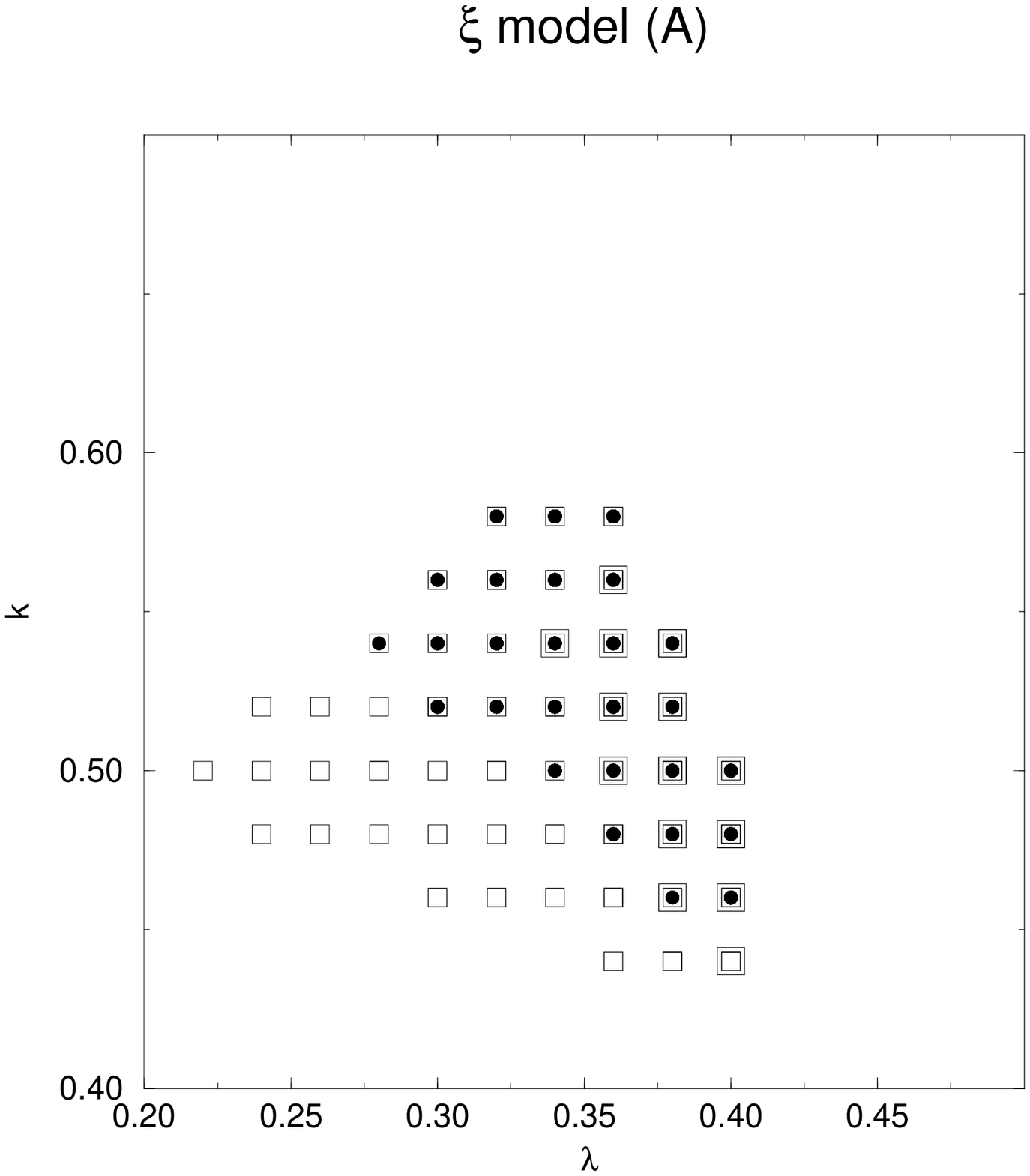}
\vspace{-0.5cm}
 }
\end{figure}
\par
\vspace{3mm}
\begin{figure}[htb]       %
\hspace{10mm}
 \parbox{60mm}{
 \epsfxsize=60mm      %
  \leavevmode
\epsfbox{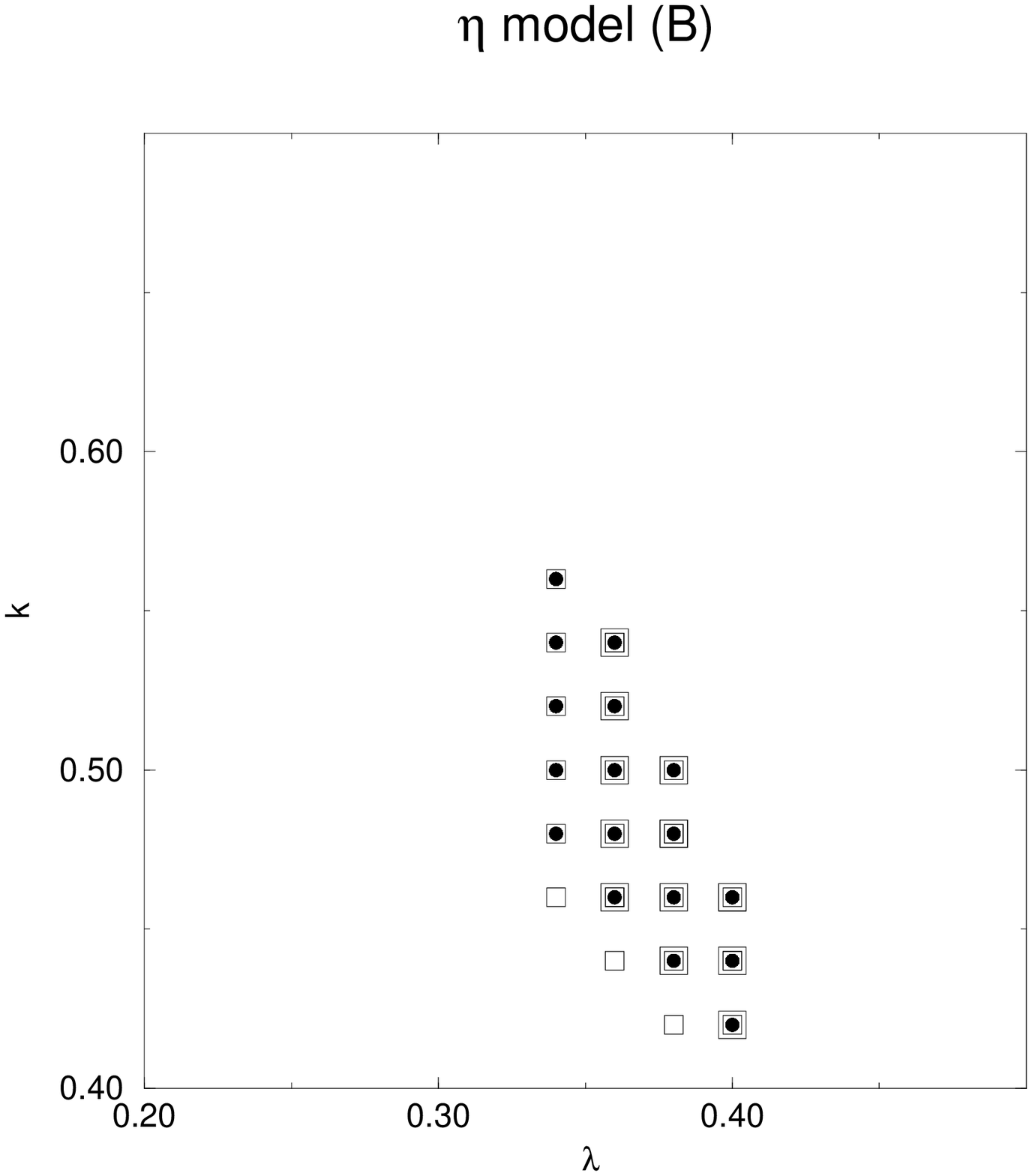}
\vspace{-0.5cm}
 }
\hspace{10mm} 
\parbox{60mm}{
 \epsfxsize=60mm      %
 \leavevmode
\epsfbox{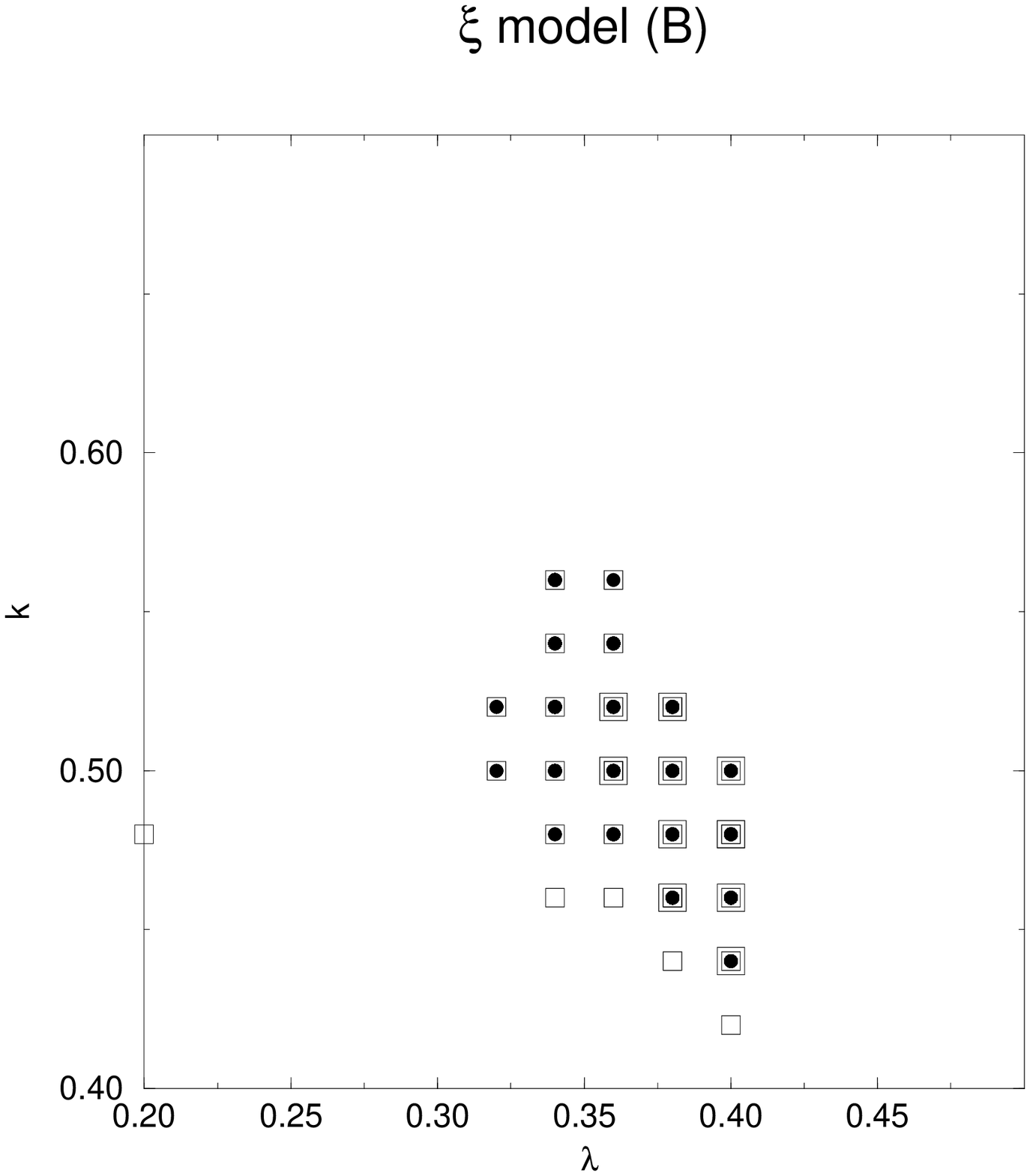}
\vspace{-0.5cm}
 }
\caption{Allowed regions of the extra U(1) models in ($\lambda$, $k$)
 plane, where (A) is for $\tan\beta=10$ and (B) is for $\tan\beta=20$.
The square satisfies both sparticle and Higgs mass constraints given 
in Eq.(17)
except for the chargino and $Z'$ boson mass bounds. 
The small circle satisfies $m_{Z'}>600$ GeV and the large
square satisfies $m_{\chi^\pm}>72$ GeV.}
\end{figure}
\end{center}

      
\section{No-scale boundary condition with non-universal gaugino masses}

In order to escape from the FCNC consideration, 
here we choose the no-scale type
boundary condition ($m_0=A_0=0$)
with the non-universal gaugino masses 
\cite{nanopoulos,komine},
\begin{eqnarray}
 0.18<M_{30}<0.36, \quad 0.25<M_{20}<1.2, \quad
    0.25<M_{Y0}=M_{X0}<1.2 \quad,
\end{eqnarray}
where we allow the non-universality among Higgs soft scalar masses in
the region $|m_i|<100$ GeV as argued in previous section.
It is well known that there is a dangerous $U(1)_{em}$ breaking
minimum due to the tachyonic slepton mass in some parameter space 
of the no-scale model. In the case of the extra U(1) models, 
the large D-term
contribution from an extra U(1) gauge multiplet is important.
In the $\eta$ model, $\tilde b_R$, $\tilde \tau_L$,
$\tilde \nu_{\tau L}$, $\tilde g$ and $\tilde g^c$ get negative 
squared mass contribution from a D-term. On the other hand, only
$\tilde g$ and $\tilde g^c$ get negative squared mass contribution
in the $\xi^-$ model. But in the both models, a right handed stau 
gets a large positive squred mass contribution unlikely 
in the case of the MSSM.
Taking account of the RGE evolution effect, the 
values of soft scalar masses at $M_S$ are given by using gaugino masses
as follows,
\begin{eqnarray}
m^2_Q &\sim& 0.26 M^2_{20} +1.90M^2_{30}-0.07M_{20}M_{30}
      -0.01M_{Y0}M_{30}-0.01M_{X0}M_{30},\nonumber \\
m^2_{\bar U} &\sim& 0.03 M^2_{Y0}-0.12 M^2_{20}+1.41 M^2_{30}
      -0.14M_{20}M_{30}-0.02M_{Y0}M_{30}-0.02M_{X0}M_{30}, \nonumber\\
m^2_{\bar D} &\sim& 0.01 M^2_{Y0}+2.37 M^2_{30},\nonumber\\
m^2_L &\sim& 0.03 M^2_{Y0}+0.33 M^2_{20},\nonumber\\
m^2_{\bar E} &\sim& 0.12 M^2_{Y0}+0.01 M^2_{X0},\nonumber\\
m^2_{\bar N} &\sim&0.08 M^2_{X0},\nonumber\\
m^2_{g^c} &\sim& 0.01 M^2_{Y0}+2.15 M^2_{30},\nonumber\\
m^2_g &\sim& 0.01 M^2_{Y0}+2.15 M^2_{30}+0.04M^2_{X0},\nonumber
\end{eqnarray} 
where we took $\tan\beta=10$, $k=0.6$ and  $\lambda=0.3$
and used the $\eta$-model RGE given in appendix B. Because of the large
gluino mass contribution there is no problem 
against the color breaking minimum,
so the color and charge conservation conditions are always satisfied 
in the $\xi^-$ model. However, it is not always the case 
for the $\eta$ model because of
the negative D-term contribution to the slepton mass. 
From the other point of view,
this might be seen as a chance for the $\eta$ model to enhance $a^{SUSY}_\mu$
by the light sneutrino.

Another problem in the no-scale model is the charged LSP \cite{komine}.
In the extra U(1) models,
since there is another serious problem the superpartners of
$S_{(1)}$ and $S_{(2)}$ are massless as noticed previously,
it is rather easy to make a LSP neutral.

Recently the experimental lower bound of Higgs boson mass is raised 
to about 113.5 GeV, this
constraint is nontrivial for the no-scale model. In order to induce 
such a large Higgs mass, we need a large stop loop contribution,
but it is difficult for the no-scale model.
In the case of the extra U(1) model the tree level lightest Higgs boson
mass at the large $\tan\beta$ limit is given by
\begin{equation}
m^2_h\sim m^2_Z (1+\frac{g^2_XQ^2_2}{g^2_Y+g^2_2})\sim 
     m^2_Z(1+Q^2_2\sin^2\theta_W),
\end{equation}
where $m_h$ becomes about 100 GeV in the $\eta$ model and the
large loop correction is not necessarily required as 
compared to the MSSM. The typical range of 
the lightest Higgs boson mass is $115\sim 120$ GeV for the $\xi$ model and
$120\sim 130$ GeV for the $\eta$ model. If we include 2-loop contributions to
the Higgs mass, they give negative contributions by few GeV and
the result of our analysis may change drastically in the $\xi$ model.

The results of numerical analysis are given in Fig.4 for
the $\eta$ model and the $\xi^-$ model. 
In the case of the no-scale model 
$m_{Z'}$ becomes significantly small and the allowed region 
\begin{center}
\begin{figure}[htb]       %
\hspace{10mm}
 \parbox{60mm}{
 \epsfxsize=60mm      %
  \leavevmode
\epsfbox{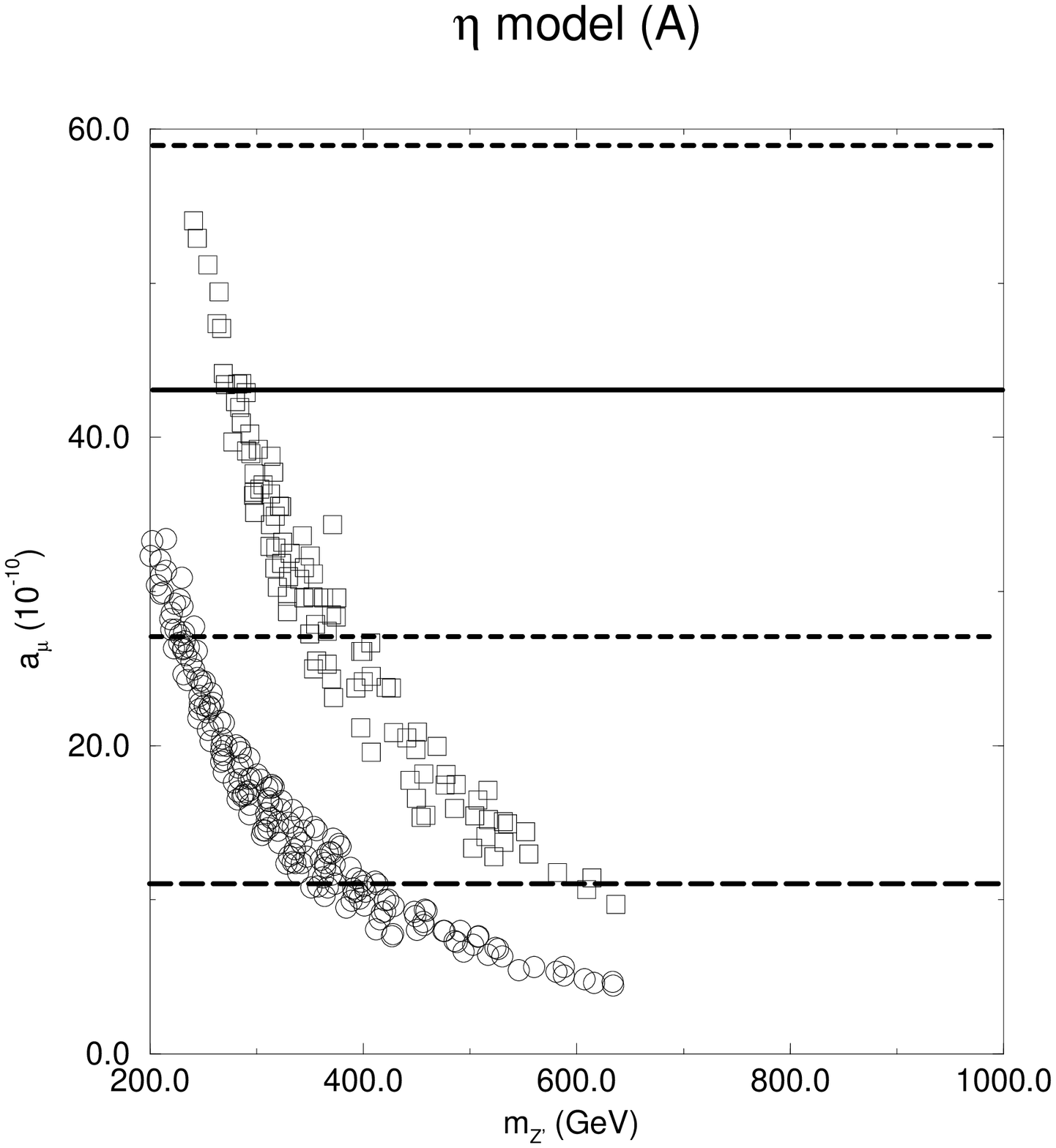}
\vspace{0.5cm}
 }
\hspace{10mm} 
\parbox{60mm}{
 \epsfxsize=60mm      %
 \leavevmode
\epsfbox{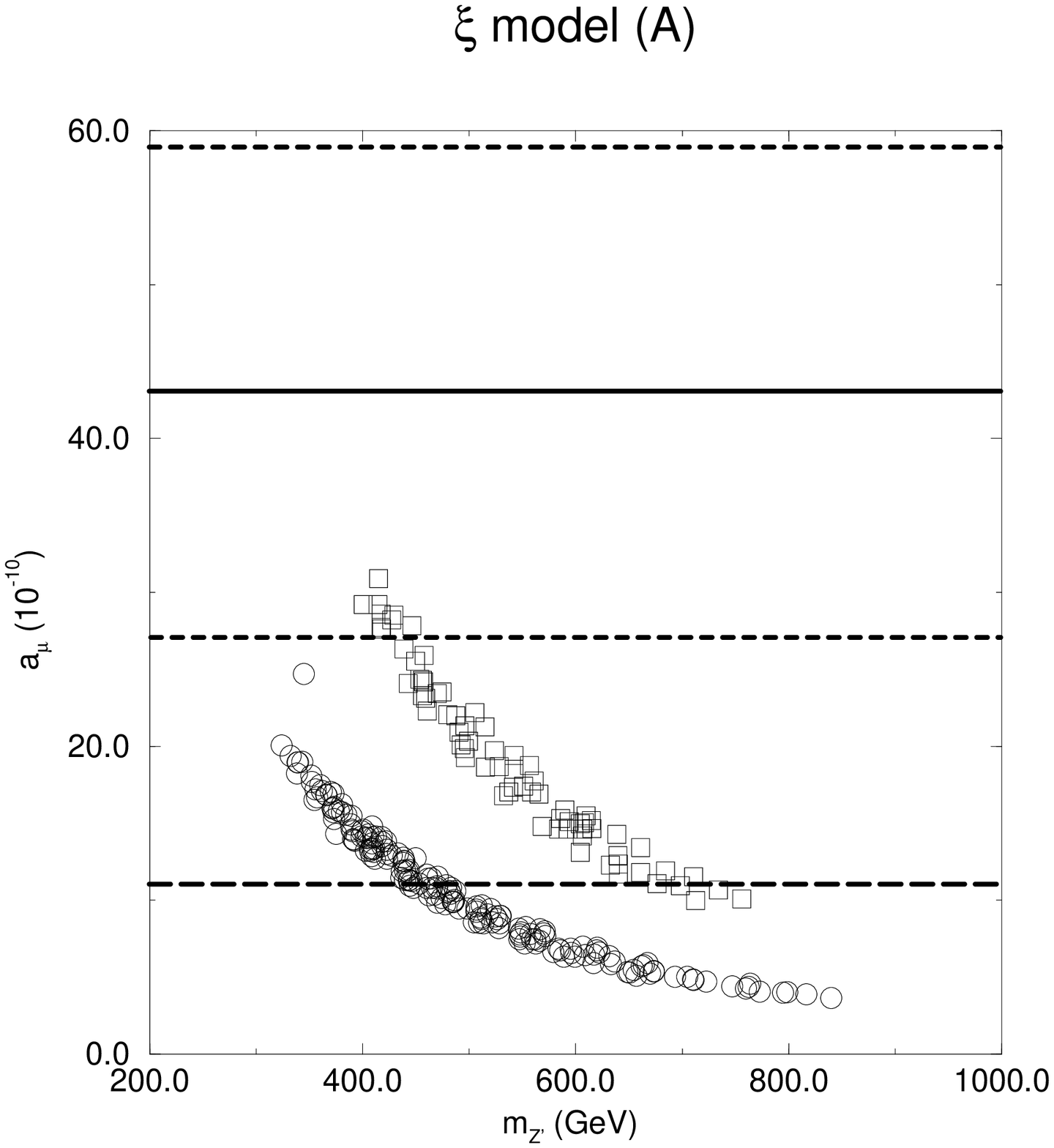}
\vspace{0.5cm}
 }
\caption{Same as Fig.2 for the no-scale model.}
\end{figure}
\end{center}
is limited in a
very narrow range in the ($a^{SUSY}_\mu$, $m_{Z'}$) plane. 
But the muon AMDM constraint never
excludes all the parameter space. As was expected, 
the values of $a^{SUSY}_\mu$
and $m_{Z'}$ are highly correlated. In order to satisfy the potential
minimum conditions, the values of $a_\lambda$ is very important.
In the large $\tan\beta$ region,     
Eq.(15) requires the small $a_\lambda$.
The $a_\lambda$ is given by
\begin{equation}
 a_\lambda=-0.02M_{Y0}-0.33M_{20}+1.34M_{30}-0.04M_{X0},
\end{equation}
which needs a large cancellation between $M_{20}$ and $M_{30}$ to 
make $a_\lambda$ small. Due to such a fine tuning structure of the
scalar potential, in the large $\tan\beta$ case, 
the phenomenologically allwed region shrinks
significantly into the ($\lambda$, $k$) plane as shown in Fig.3.
In this way $M_{20}$ is never used for tuning of the sneutrino mass,
the enhancement of $a^{SUSY}_\mu$ 
due to the small sneutrino mass does not occur.
So $a^{SUSY}_\mu$ depends only on the 
typical supersymmetry breaking scale i.e.
$m_{Z'}.$\footnote{In the extra U(1) models, since two parameters $\mu$ and 
$m_{\tilde\nu}$ in $a^{\chi^\pm}_\mu$ are strongly correlated to
$m_{Z'}$, it is expected that the same results are obtained in more general
non-universal case, because it is difficult to realize 
$\mu \ll m_{Z'}$ and $m_{\tilde\nu}\ll m_{Z'}$ simultaneously.}
As shown in Fig.4, both the $Z'$ mass bound
$m_{Z'}>600$ GeV and the muon AMDM 2$\sigma$ bound 
are satisfied only in a very narrow range of the
($a_\mu$, $m_{Z'}$) plane for $\tan\beta=20$.

Since the no-scale condition is too strong to satisfy the experimental
bound of $m_{Z'}$, we allow to add the
universal scalar mass ($m_0=200,400$ GeV) without asking its origin.
Here we allow the non-universality of Higgs soft scalar masses as
$|m_i-m_0|<50$ GeV.
In this case $m_{Z'}$ becomes large enough but the muon AMDM 1$\sigma$ bound
excludes 
almost all parameter region that has been allowed if we would not
take account of this new constraint.  
It is obvious from Fig.4 and Fig.5 that the larger the $m_0$ 
is the weaker the 
\begin{center}
\begin{figure}[htb]       %
\hspace{10mm}
 \parbox{60mm}{
 \epsfxsize=60mm      %
  \leavevmode
\epsfbox{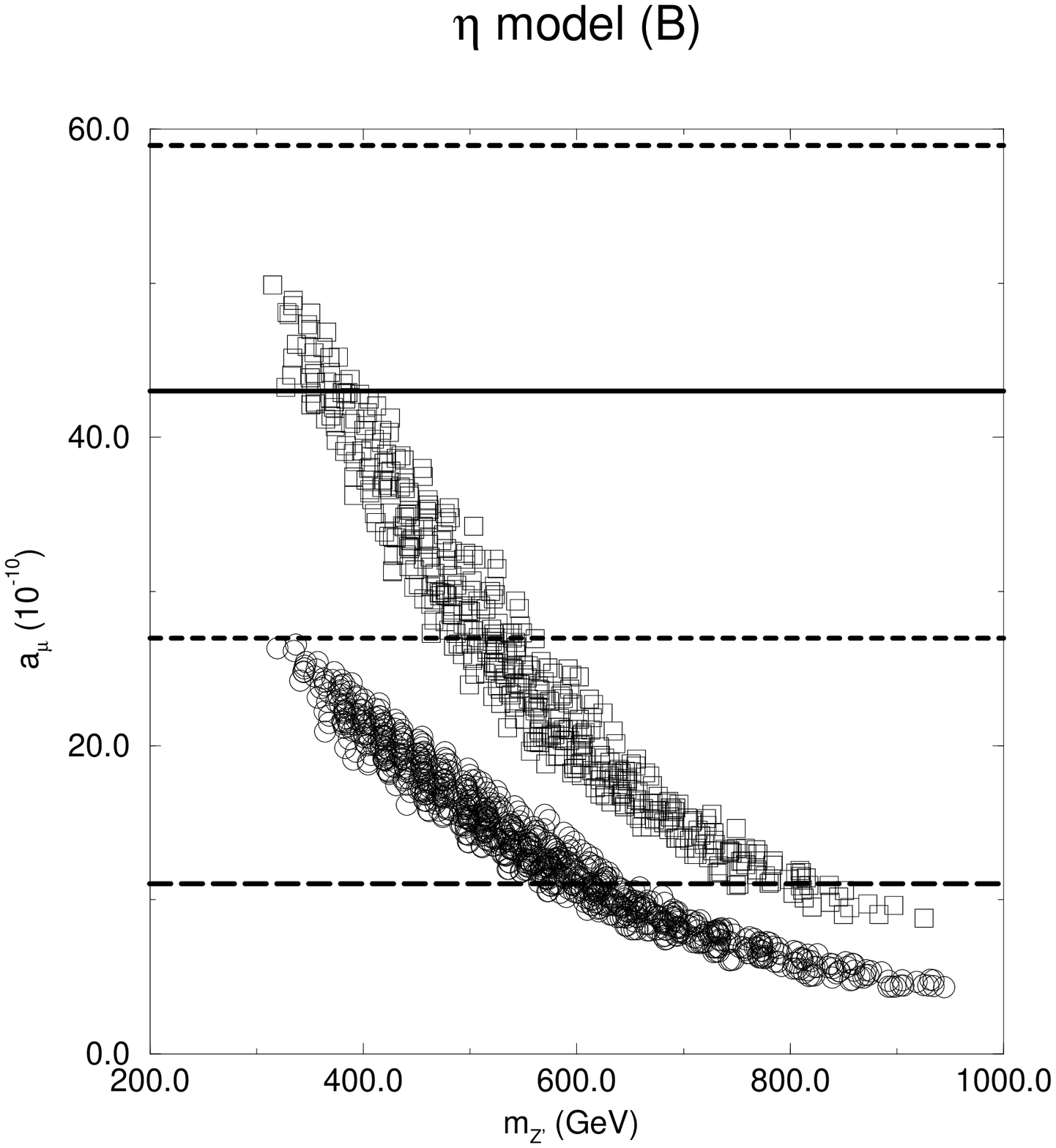}
\vspace{-0.5cm}
 }
\hspace{10mm} 
\parbox{60mm}{
 \epsfxsize=60mm      %
 \leavevmode
\epsfbox{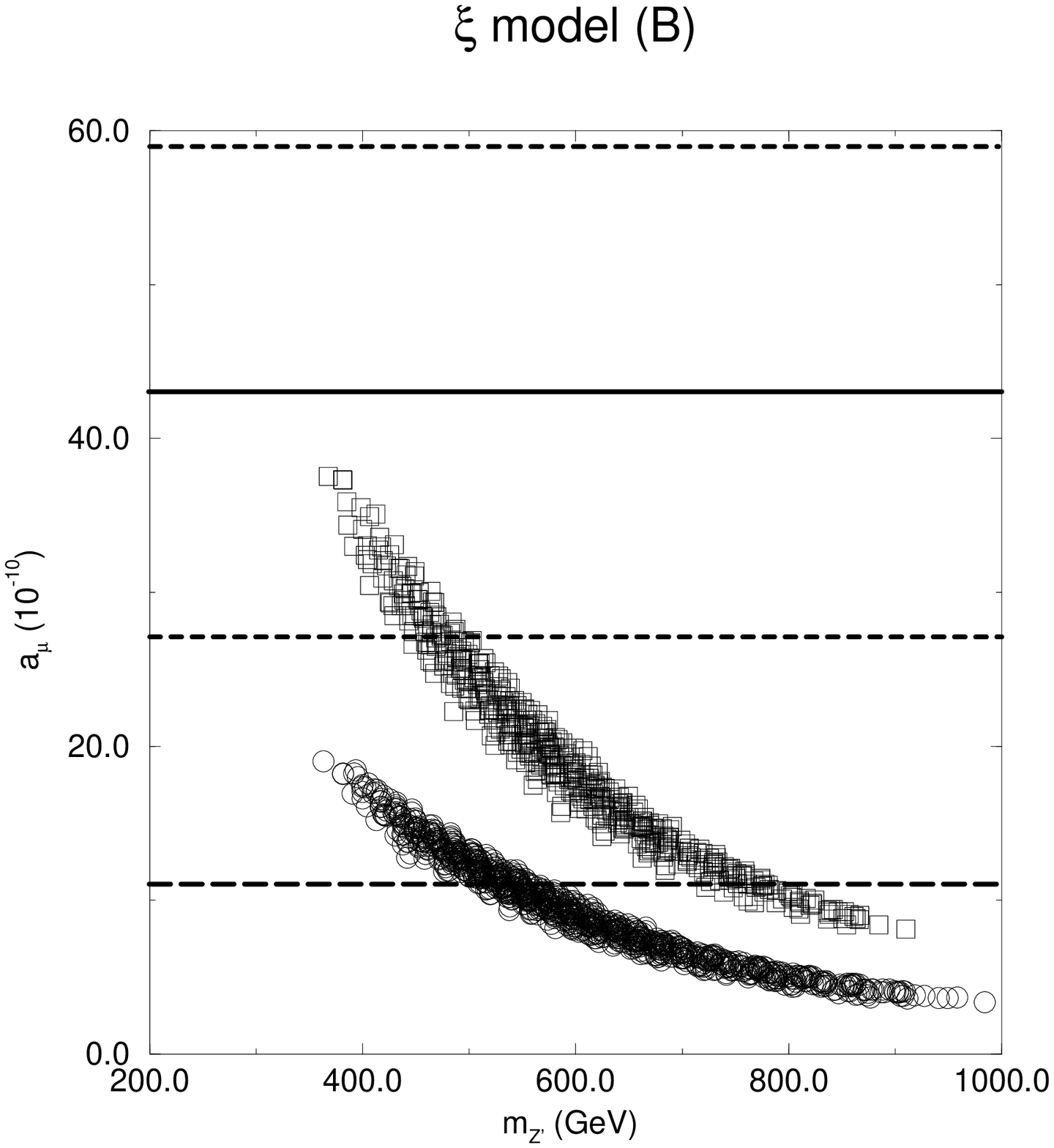}
\vspace{-0.5cm}
 }
\end{figure}
\par
\vspace{3mm}
\begin{figure}[htb]       %
\hspace{10mm}
 \parbox{60mm}{
 \epsfxsize=60mm      %
  \leavevmode
\epsfbox{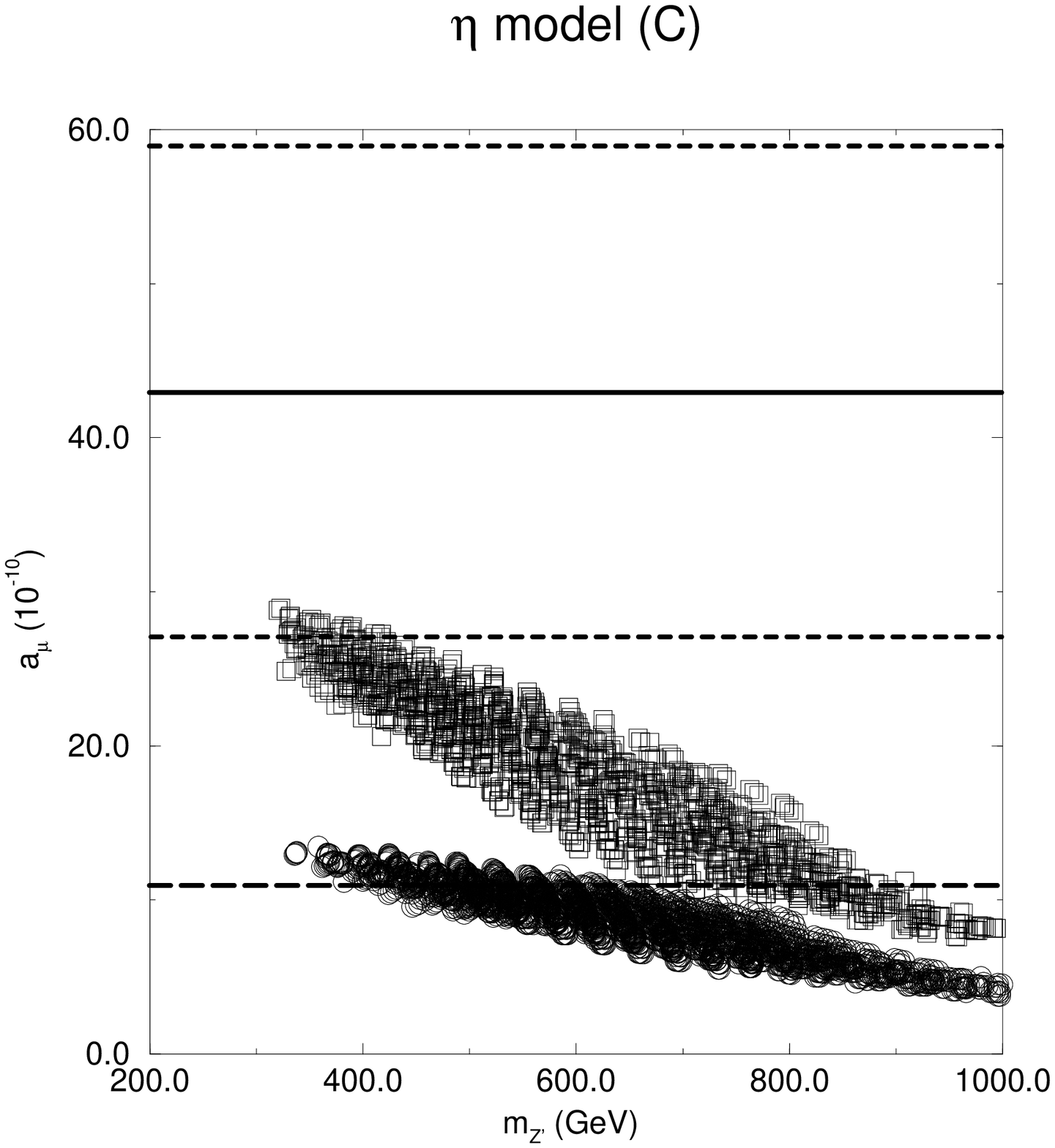}
\vspace{0.5cm}
 }
\hspace{10mm} 
\parbox{60mm}{
 \epsfxsize=60mm      %
 \leavevmode
\epsfbox{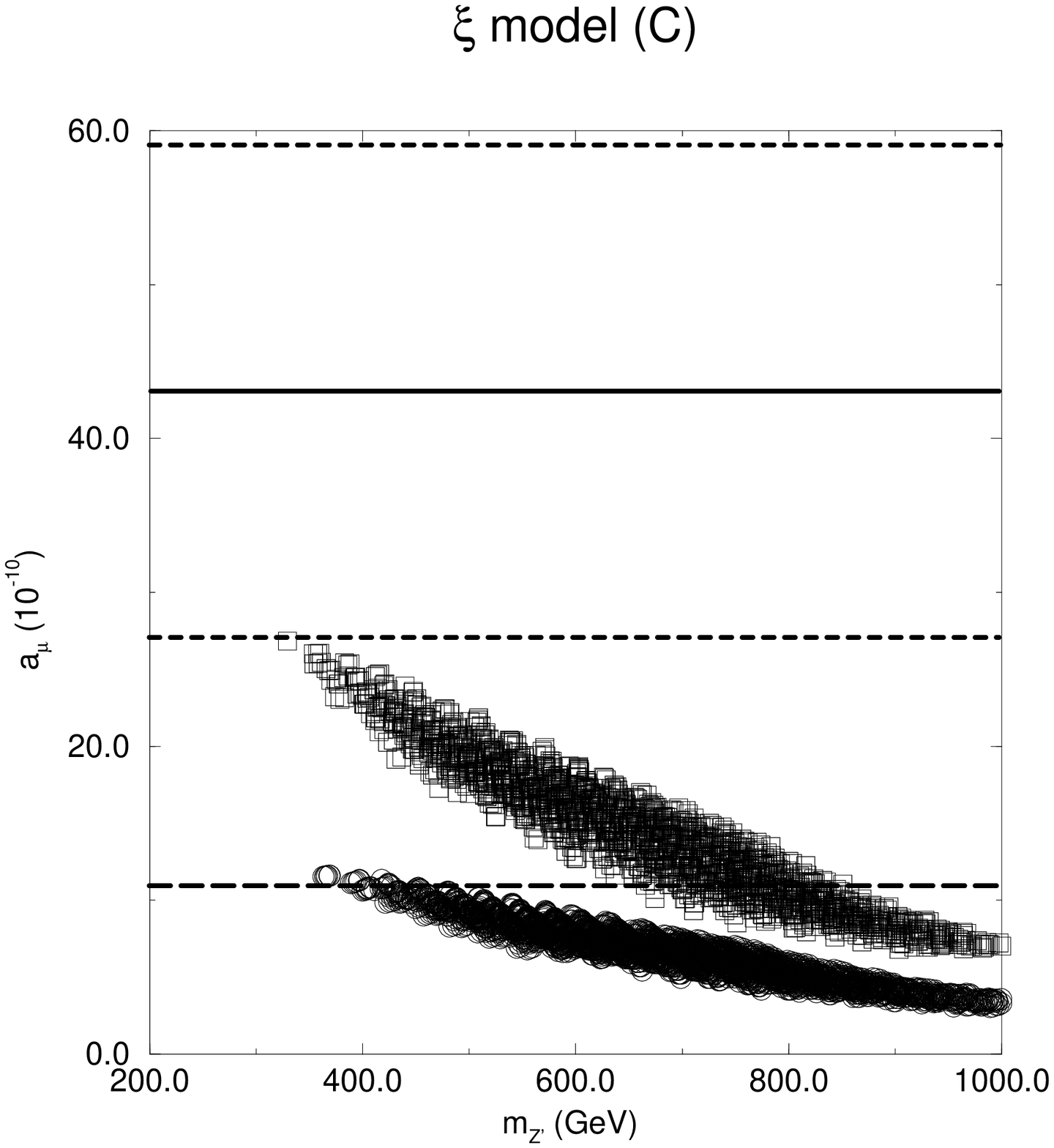}
\vspace{0.5cm}
 }
\caption{The same ones as Fig.3 except for (B) is for
$m_0=200$ GeV and (C) is for $m_0=400$ GeV.}
\end{figure}
\end{center}   
correlation between $a_\mu$ and $m_{Z'}$ is. 
The reason is that the soft scalar mass domiates
over the D-term contributions
in the mass formula of sneutrino (see appendix C).
The dominating soft scalar mass weakens the correlation between $m_{Z'}$
and $m_{\tilde\nu}$.
For $\tan\beta=20$, the $2\sigma$ bound gives the upper
bound of extra neutral boson mass ($m^{2\sigma}_{Z'}$)
about $600$ (A), $800$ (B) and
$900$ GeV (C) in the $\eta$ model and about 
$700$ (A), $800$ (B) and $850$ GeV (C) in the $\xi$ model.
Although it is shown that the larger $m_0$ makes $m^{2\sigma}_{Z'}$
larger, the value of $m^{2\sigma}_{Z'}$
seems to be saturated around $900$ GeV 
in both models.

\section{Summary}

We have estimated the allowed region of the extra neutral boson mass
in taking account of the new measurement of the 
muon AMDM in the $\mu$-problem
solvable extra U(1) model. We focussed our attention only on 
the chargino-sneutrino
and neutralino-smuon contributions to $a^{SUSY}_\mu$. Another 
exotic contribution from leptoquark exchange was neglected. 
Although such a loop
contribution needs the enhancement
either due to the large $\tan\beta$ 
or the light sneutrino as in the case of the MSSM
in order to explain the muon AMDM data, 
the allowed parameter region of the extra U(1) model shrinks
significantly as compared to the MSSM.
The reason is that
the large $\tan\beta$ solution is allowed only in a very narrow region 
of the ($\lambda$, $k$) plane and the light sneutrino solution
contradicts the phenomenological constraint on the extra neutral gauge
boson mass or the chargino mass.
 
In the case of the minimal supergravity,
in order to preserve the hierarchy such as $u\gg v_{1,2}$ and 
$v_2\gg v_1$ against the quantum correction, the light gluino
solution is favored. Due to  the gaugino mass universality,
this requires lighter chargino so that the chargino mass lower bound is
more stringent. 
Because of these obstacles, the minimul supergravity scenario 
is excluded by the muon AMDM constraint. 

In the case of the non-universal gaugino
mass, the chargino mass constraint disappears in the large $M_{20}$ region.
Thus the small window is still remained
for a no-scale model. 
Because of the strong correlation between $m_{Z'}$ and $a^{SUSY}_\mu$ in
the extra U(1) models, 
we can expect to get a new information on the upper bound of $m_{Z'}$ from 
the further improvement of the $(g-2)_\mu$ measurement. \\

{\em Acknowledgements}: The authors thanks D.Suematsu for helpful
discussions.

\appendix

\section{Notations}

In this paper, we use the following superpotential:
\begin{eqnarray}
W&=& -y_t(t_Lt_R H^0_2-t_Rb_LH^+_2)-y_b(b_Lb_R H^0_1-b_Rt_LH^-_1)
     -kSg\bar{g} \nonumber \\
 &-&y_\tau(\tau_L\tau_R H^0_1-\tau_R\nu_LH^-_1)
     +\lambda S(H^0_1H^0_2-H^-_1H^+_2),
\end{eqnarray}
and the soft SUSY breaking terms:
\begin{eqnarray}
{\cal L}_{soft} &=&\int d\theta^2 \theta^2(a W 
      +\frac12 M W^\alpha W_\alpha) \nonumber \\
   &=& \frac12 M_Y\lambda_Y\lambda_Y
       +\frac12 M_2\lambda^{(a)}_2\lambda^{(a)}_2
       +\frac12 M_X\lambda_X\lambda_X \nonumber \\
   &-& y_t a_t (H^0_2\tilde{t_R}\tilde{t_L}-H^+_2\tilde{t_R}\tilde{b_L})
      -y_b a_b (H^0_1\tilde{b_R}\tilde{b_L}-H^-_1\tilde{b_R}\tilde{t_L}) 
      \nonumber\\
  &-& y_\tau a_\tau (H^0_1\tilde{\tau_R}\tilde{\tau_L}
       -H^-_1\tilde{\tau_R}\tilde{\nu_\tau})
     +\lambda a_\lambda S(H^0_1H^0_2-H^-_1H^+_2) \nonumber \\
      &-& ka_kS\tilde{g}\tilde{g^c} +h.c..
\end{eqnarray}
In this notation, any term in the RGEs of $a_i$ apper 
with same sign\cite{martin, yamada}. 
The muon-chargino and the muon-neutralino interactions are described by
\begin{eqnarray}
{\cal L}_\mu &=& y_\mu[\mu_L\tilde\mu_R\tilde H^0_1
                   +\mu_R(\tilde H^0_1\tilde\mu_L-\tilde\nu_L\tilde H^-_1)]
                \nonumber \\
         &-&\frac{1}{\sqrt{2}}(\tilde\nu^*_L,\tilde\mu^*_L)
	  \left(
                           \begin{array}{cc}
                       g_2\lambda^{(3)}_2-g_Y\lambda_Y+2g_XQ_L\lambda_X
                             & g_2(\lambda^{(1)}_2-i\lambda^{(2)}_2) \\
                               g_2(\lambda^{(1)}_2+i\lambda^{(2)}_2)
                       & -g_2\lambda^{(3)}_2-g_Y\lambda_Y+2g_XQ_L\lambda_X \\
                           \end{array}
                          \right) 
         \left(
	   \begin{array}{c}
	    \nu_L \\ \mu_L
	   \end{array}
         \right) \nonumber \\
      &-&\sqrt{2}\mu_R(g_Y\lambda_Y+g_XQ_E\lambda_X)\tilde\mu^*_R+h.c.
\end{eqnarray}
and the chargino and the neutralino mass terms are given by
\begin{eqnarray}
 {\cal L}_M &=& -\lambda [u(\tilde H^0_1\tilde H^0_2
                         -\tilde H^-_1\tilde H^+_2)
                   +v_1\tilde S\tilde H^0_2 +v_2\tilde S\tilde H^0_1]
              \nonumber \\
       &-&\frac{1}{\sqrt{2}}(v_1,0)
	  \left(
                           \begin{array}{cc}
                       g_2\lambda^{(3)}_2-g_Y\lambda_Y+2g_XQ_L\lambda_X
                             & g_2(\lambda^{(1)}_2-i\lambda^{(2)}_2) \\
                               g_2(\lambda^{(1)}_2+i\lambda^{(2)}_2)
                       & -g_2\lambda^{(3)}_2-g_Y\lambda_Y+2g_XQ_L\lambda_X \\
                           \end{array}
                          \right) 
         \left(
	   \begin{array}{c}
	    \tilde H^0_1 \\ \tilde H^-_1
	   \end{array}
         \right) \nonumber \\
       &-&\frac{1}{\sqrt{2}}(0,v_2)
	  \left(
                           \begin{array}{cc}
                       g_2\lambda^{(3)}_2+g_Y\lambda_Y+2g_XQ_L\lambda_X
                             & g_2(\lambda^{(1)}_2-i\lambda^{(2)}_2) \\
                               g_2(\lambda^{(1)}_2+i\lambda^{(2)}_2)
                       & -g_2\lambda^{(3)}_2+g_Y\lambda_Y+2g_XQ_L\lambda_X \\
                           \end{array}
                          \right) 
         \left(
	   \begin{array}{c}
	    \tilde H^+_2 \\ \tilde H^0_2
	   \end{array}
         \right) \nonumber \\
      &-&\sqrt{2}\tilde S(g_XQ_S\lambda_X)S^*
       +\frac12 M_Y\lambda_Y\lambda_Y
        +\frac12 M_2\lambda^{(a)}_2\lambda^{(a)}_2+h.c. \nonumber \\
      &=& -\frac12 \chi^{0T}
       \left(
	\begin{array}{cccccc}
	 -M_Y & 0 & 0 &-\frac{g_Yv_1}{\sqrt{2}}&\frac{g_Yv_2}{\sqrt{2}}& 0 \\
         0 & -M_2 & 0 &\frac{g_2v_1}{\sqrt{2}}&-\frac{g_2v_2}{\sqrt{2}}& 0 \\
         0 & 0 & -M_X & \sqrt{2}g_XQ_{H_1}v_1 & \sqrt{2}g_XQ_{H_2}v_2 &
          \sqrt{2}g_XQ_S u \\
	 -\frac{g_Yv_1}{\sqrt{2}}& \frac{g_2v_1}{\sqrt{2}} &
	   \sqrt{2}g_XQ_{H_1}v_1 & 0 &\lambda u& \lambda v_2 \\
	  \frac{g_Yv_2}{\sqrt{2}} & -\frac{g_2v_2}{\sqrt{2}}&
	   \sqrt{2}g_XQ_{H_2}v_2 &\lambda u & 0 & \lambda v_1 \\
	 0&0& \sqrt{2}g_XQ_S u &\lambda v_2 & \lambda v_1 & 0
	\end{array}
       \right) \chi^0 \nonumber \\
        &+& (\tilde w^+,\tilde H^+_2)
        \left(
	 \begin{array}{cc}
	    M_2 & g_2v_1 \\
	    g_2v_2 & \lambda u
	 \end{array}
	\right) 
	\left(
	 \begin{array}{c}
	  \tilde w^- \\ \tilde H^-_1
	 \end{array}
	 \right)+h.c. 
\end{eqnarray}
where the six components of the neutralino $\chi^0$ 
and the chargino $\tilde w^\pm$ are
defined as
\begin{eqnarray}
 \chi^{0T}&=&(\begin{array}{cccccc}
	  \lambda_Y, & \lambda^{(3)}_2, & \lambda_X, & \tilde H^0_1, &
	  \tilde H^0_2, & \tilde S
	\end{array}), \nonumber \\
 \tilde w^\pm &=& -\frac{\lambda^{(1)}_2\mp i\lambda^{(2)}_2}{\sqrt{2}}.
\end{eqnarray}
More general case in which the neutralino mass matrix includes the
U(1) gauge kinetic term mixing effect is argued in \cite{daijiro}. 
 Interaction terms are
\begin{eqnarray}
{\cal L}_C&=&\bar\mu [g_2P_R\tilde w^+\tilde\nu^*
        -y_\mu P_L\tilde H^-_1\tilde\nu] +h.c.,\\
{\cal L}_N&=&\bar\mu [P_R\{y_\mu\tilde\mu_R\tilde H^0_1
        +\frac{1}{\sqrt{2}}\tilde\mu^*_L (g_2\lambda^{(3)}_2
        +g_Y\lambda_Y-2g_XQ_L\lambda_X)\} \nonumber \\
        &+&P_L\{y_\mu\tilde H^0_1\tilde\mu_L-\sqrt{2}\tilde\mu_R
          (g_Y\lambda_Y+g_XQ_E\lambda_X)\}]+h.c.,
\end{eqnarray}
where $P_L=(1-\gamma_5)/2$ and $P_R=(1+\gamma_5)/2$.
We can diagonarize the mass matrices $M_{\chi^\pm}$ ,$M_{\chi^0}$
and $M^2_{\tilde \mu}$ by unitary matrix $U_{\chi^\pm}$, $U_{\chi^0}$
and $U_{\tilde\mu}$ as
\begin{eqnarray}
(U^\dagger_{\chi^0}M_{\chi^0}U_{\chi^0})&=&m_{\chi^0X}\delta_{XY}
 \quad (X,Y=1-6), \\
(U^\dagger_{\chi^+}M_{\chi^\pm}U_{\chi^-})&=&m_{\chi^\pm X}\delta_{XY}
 \quad (X,Y=1,2), \\
(U^\dagger_{\tilde\mu}M^2_{\tilde\mu}U_{\tilde\mu})
    &=&m^2_{\tilde\mu A}\delta_{AB}
 \quad (A,B=1,2), 
\end{eqnarray}
respectively. 
In this base the muon-chargino and the muon-neutralino interaction terms are
rewritten as
\begin{eqnarray}
{\cal L}_{int}&=&\sum_{AX}\bar\mu (N^L_{AX}P_L+N^R_{AX}P_R)
     \chi^0_X\tilde\mu_A \nonumber \\
  &+& \sum_X \bar\mu (C^L_XP_L+C^R_XP_R)\chi^\pm_X\tilde\nu+h.c.,\\
 C^L_X&=&y_\mu(U_{\chi^-})_{2X}, \\
 C^R_X&=&-g_2(U_{\chi^+})_{1X}, \\
 N^L_{AX}&=&-y_\mu(U_{\chi^0})_{4X}(U_{\tilde\mu})_{LA}
  -\sqrt{2}[g_Y(U_{\chi^0})_{1X}+Q_E g_X(U_{\chi^0})_{3X}]
       (U_{\tilde\mu})_{RA},\\
 N^R_{AX}&=&[\frac{g_2}{\sqrt{2}}(U_{\chi^0})_{2X}
      +\frac{g_Y}{\sqrt{2}}(U_{\chi^0})_{1X}
      -\sqrt{2}Q_Lg_X(U_{\chi^0})_{3X}](U_{\tilde\mu})_{LA} \nonumber \\
   &-&y_\mu(U_{\chi^0})_{4X}(U_{\tilde\mu})_{RA},
\end{eqnarray}
where we redefined the superfield as $H^-_1\to -H^-_1$ and $\mu_L\to -\mu_L$
to use the same notation as \cite{moroi} except for smuon mass mixing. 

The neutralino-smuon loop contribution is 
\begin{eqnarray}
{\Delta a^{\chi^0}_\mu} &=&\frac{m_\mu}{16\pi^2}
    \sum_{AX}[
	      -\frac{m_\mu}{6m^2_{\tilde\mu A}(1-x_{AX})^4}
              (|N^L_{AX}|^2+|N^R_{AX}|^2) \nonumber \\
      &\times& (1-6x_{AX}+3x^2_{AX}+2x^3_{AX}-6x^2_{AX}\log x_{AX}) 
             \nonumber \\
       &-& \frac{m_{\chi^0 X}}{m^2_{\tilde\mu A}(1-x_{AX})^3}
           N^L_{AX}N^R_{AX}(1-x^2_{AX}+2x_{AX}\log x_{AX})
              ],\\
 x_{AX}&=&\frac{m^2_{\chi^0 X}}{m^2_{\tilde\mu A}}, \nonumber
\end{eqnarray}
and the chargino-sneutrino loop contribution is
\begin{eqnarray}
\Delta a^{\chi^\pm}_\mu &=&\frac{m_\mu}{16\pi^2}
 \sum_X [
	\frac{m_\mu}{3m^2_{\tilde\nu}(1-x_{X})^4}
	(|C^L_X|^2+|C^R_X|^2) \nonumber \\
     &\times& (1+\frac32x_X-3x^2_X+\frac12x^3_X+3x^2_X\log x_X) \nonumber \\
 &-& \frac{3m_{\chi^\mp X}}{m^2_{\tilde\nu}(1-x_X)^3}
     C^L_XC^R_X(1-\frac43x_X+\frac13 x^2_X+\frac23\log x_X)
       ], \\
 x_X&=&\frac{m^2_{\chi^\pm X}}{m^2_{\tilde\nu}}. \nonumber
\end{eqnarray}

\section{The RGEs}

In our notation, the renormalization group equations of soft breaking
terms in the $\eta$ model are given by
\begin{eqnarray}
{(2\pi)}\frac{a_t}{dt}&=& 6Y_t a_t+Y_b a_b+Y_\lambda a_\lambda
         +\frac{16}{3}\alpha_3 M_3 +3\alpha_2 M_2+\frac{13}{9}\alpha_Y M_Y 
         +\frac43\alpha_X M_X, \nonumber \\
{(2\pi)}\frac{a_b}{dt}&=& 6Y_b a_b+Y_t a_t+Y_\tau a_\tau +Y_\lambda a_\lambda
         +\frac{16}{3}\alpha_3 M_3 +3\alpha_2 M_2+\frac79 \alpha_Y M_Y 
         +\frac13\alpha_X M_X,  \nonumber\\
{(2\pi)}\frac{a_\tau}{dt}&=& 4Y_\tau a_\tau+3Y_b a_b+Y_\lambda a_\lambda
         +3\alpha_2 M_2+3\alpha_Y M_Y +\frac13\alpha_X M_X,  \nonumber\\
{(2\pi)}\frac{a_\lambda}{dt}&=&3Y_t a_t+3Y_b a_b+Y_\tau a_\tau+9Y_k a_k
        +4Y_\lambda a_\lambda
        +3\alpha_2 M_2+\alpha_Y M_Y 
        +\frac73\alpha_X M_X,  \nonumber\\
{(2\pi)}\frac{a_k}{dt}&=&11Y_k a_k+2Y_\lambda a_\lambda
        +\frac{16}{3}\alpha_3 M_3 +\frac49\alpha_Y M_Y +\frac73\alpha_X M_X,
	\nonumber
\end{eqnarray}
where $\alpha=\frac{g^2}{4\pi}$ and $Y=\frac{y^2}{4\pi}$ .
For the soft scalar masses they are expressed as 
\begin{eqnarray}
{(2\pi)}\frac{m^2_{Q_{(3)}}}{dt} &=& Y_t M^2_t +Y_b M^2_b -3\alpha_2 M^2_2
         -\frac{16}{3}\alpha_3M^2_3 -\frac19\alpha_Y M^2_Y
         -\frac49\alpha_X M^2_X, \nonumber\\
{(2\pi)}\frac{m^2_{\bar U_{(3)}}}{dt} &=& 2Y_t M^2_t 
         -\frac{16}{3}\alpha_3M^2_3 -\frac{16}{9}\alpha_Y M^2_Y
         -\frac49\alpha_X M^2_X, \nonumber\\
{(2\pi)}\frac{m^2_{\bar D_{(3)}}}{dt} &=& 2Y_b M^2_b 
         -\frac{16}{3}\alpha_3M^2_3 -\frac49\alpha_Y M^2_Y
         -\frac19\alpha_X M^2_X, \nonumber\\
{(2\pi)}\frac{m^2_{L_3}}{dt} &=& Y_\tau M^2_\tau -3\alpha_2 M^2_2
         -\alpha_Y M^2_Y -\frac19\alpha_X M^2_X, \nonumber\\
{(2\pi)}\frac{m^2_{\bar E_{(3)}}}{dt} &=& 2Y_\tau M^2_\tau 
         -4\alpha_Y M^2_Y
         -\frac49\alpha_X M^2_X, \nonumber\\
{(2\pi)}\frac{m^2_{\bar N_{(i)}}}{dt} &=& 
          -\frac{25}{9}\alpha_X M^2_X, \nonumber\\
{(2\pi)}\frac{m^2_{H_{1(3)}}}{dt} &=& 3Y_b M^2_b 
         +Y_\lambda M^2_\lambda -3\alpha_2 M^2_2-\alpha_Y M^2_Y
         -\frac19\alpha_X M^2_X, \nonumber\\
{(2\pi)}\frac{m^2_{H_{2(3)}}}{dt} &=& 3Y_t M^2_t +Y_\tau M^2_\tau 
         +Y_\lambda M^2_\lambda -3\alpha_2 M^2_2-\alpha_Y M^2_Y
         -\frac{16}{9}\alpha_X M^2_X, \nonumber\\
{(2\pi)}\frac{m^2_{g_{(i)}}}{dt} &=& Y_k M^2_k -\frac49\alpha_Y M^2_Y
         -\frac{16}{3}\alpha_3M^2_3
         -\frac{16}{9}\alpha_X M^2_X, \nonumber\\
{(2\pi)}\frac{m^2_{g^c_{(i)}}}{dt} &=& Y_k M^2_k -\frac49\alpha_Y M^2_Y
         -\frac{16}{3}\alpha_3M^2_3
         -\frac19\alpha_X M^2_X, \nonumber\\
{(2\pi)}\frac{m^2_{S_{(3)}}}{dt} &=& 9Y_k M^2_k +2Y_\lambda M^2_\lambda
         -\frac{25}{9}\alpha_X M^2_X, \nonumber
\end{eqnarray}
where we omitted the two-loop contributions, for simplicity, and 
\begin{eqnarray}
M^2_t &=& m^2_{Q_{(3)}}+m^2_{\bar U_{(3)}}+m^2_{H_{2(3)}}+a^2_t,\nonumber\\
M^2_b &=& m^2_{Q_{(3)}}+m^2_{\bar D_{(3)}}+m^2_{H_{1(3)}}+a^2_b,\nonumber\\
M^2_\tau &=&m^2_{L_{(3)}}+m^2_{\bar E_{(3)}}+m^2_{H_{1(3)}}+a^2_\tau,
      \nonumber\\
M^2_\lambda &=&m^2_{H_{1(3)}}+m^2_{H_{2(3)}}+m^2_{S_{(3)}}
      +a^2_\lambda,\nonumber\\
M^2_k &=&m^2_g+m^2_{\bar g}+m^2_{S_{(3)}}+a^2_k, \nonumber
\end{eqnarray}
and $m^2_{g{(1)}}=m^2_{g{(2)}}=m^2_{g{(3)}}=m^2_g$ is assumed.

\section{Sfermion spectrums}
In our notation, a sfermion mass matrix is given by 
\begin{eqnarray}
M^2_{\tilde t}&=&\left(
                           \begin{array}{cc}
                        m^2_{\tilde t_L}+y^2_t v^2_2 
                             & y_tv_2(a_t-\lambda u\cot\beta) \\
                               y_tv_2(a_t-\lambda u\cot\beta)
                        & m^2_{\tilde t_R} +y^2_tv^2_2  \\
                           \end{array}
                          \right) , \nonumber \\
M^2_{\tilde b}&=&\left(
                           \begin{array}{cc}
                        m^2_{\tilde b_L}+y^2_b v^2_1 
                             & y_bv_1(a_b-\lambda u\tan\beta) \\
                               y_bv_1(a_b-\lambda u\tan\beta)
                        & m^2_{\tilde b_R} +y^2_bv^2_1  \\
                           \end{array}
                          \right) , \nonumber \\
M^2_{\tilde \tau}&=&\left(
                           \begin{array}{cc}
                        m^2_{\tilde\tau_L}+y^2_t v^2_2 
                             & y_\tau v_1(a_\tau-\lambda u\tan\beta) \\
                               y_\tau v_1(a_\tau-\lambda u\tan\beta)
                        & m^2_{\tilde\tau_R} +y^2_\tau v^2_1  \\
                           \end{array}
                          \right) , \nonumber \\
M^2_{\tilde g}&=&\left(
                           \begin{array}{cc}
                        m^2_{\tilde g}+k^2 u^2 
                             & a_kku-\lambda kv_1v_2 \\
                               a_kku-\lambda kv_1v_2
                        & m^2_{\tilde g^c} +k^2u^2  \\
                           \end{array}
                          \right) , \nonumber \\
M^2_{\tilde\nu_\tau}&=&m^2_{\tilde\nu_\tau}, \nonumber
\end{eqnarray}
where
\begin{eqnarray}
m^2_{\tilde t_L}&=& m^2_{Q_3}+\frac{3g^2_2-g^2_Y}{12}(v^2_1-v^2_2)
       -\frac13 g^2_X(\frac16 v^2_1+\frac23 v^2_2-\frac56  u^2)
   , \nonumber\\
m^2_{\tilde t_R}&=& m^2_{\bar U_3}+\frac{g^2_Y}{3}(v^2_1-v^2_2)
       -\frac13 g^2_X(\frac16 v^2_1+\frac23 v^2_2-\frac56  u^2)
   , \nonumber\\
m^2_{\tilde b_L}&=& m^2_{Q_3}-\frac{3g^2_2+g^2_Y}{12}(v^2_1-v^2_2)
       -\frac13 g^2_X(\frac16 v^2_1+\frac23 v^2_2-\frac56  u^2)
   , \nonumber\\
m^2_{\tilde b_R}&=& m^2_{\bar D_3}-\frac{g^2_Y}{6}(v^2_1-v^2_2)
       +\frac16 g^2_X(\frac16 v^2_1+\frac23 v^2_2-\frac56  u^2)
   , \nonumber\\
m^2_{\tilde \tau_L}&=& m^2_{L_3}+\frac{-g^2_2+g^2_Y}{4}(v^2_1-v^2_2)
       +\frac16 g^2_X(\frac16 v^2_1+\frac23 v^2_2-\frac56  u^2)
   , \nonumber\\
m^2_{\tilde \tau_R}&=& m^2_{\bar E_3}-\frac{g^2_Y}{2}(v^2_1-v^2_2)
       -\frac13 g^2_X(\frac16 v^2_1+\frac23 v^2_2-\frac56  u^2) 
   , \nonumber\\
m^2_{\tilde\nu_\tau}&=& m^2_{L_3}+\frac{g^2_2+g^2_Y}{4}(v^2_1-v^2_2)
      +\frac16 g^2_X(\frac16 v^2_1+\frac23 v^2_2-\frac56  u^2)
   , \nonumber\\
m^2_{\tilde g} &=& m^2_g +\frac{g^2_Y}{6}(v^2_1-v^2_2)
       +\frac23 g^2_X (\frac16 v^2_1+\frac23 v^2_2-\frac56  u^2)
   , \nonumber\\
m^2_{\tilde g^c} &=& m^2_{g^c} -\frac{g^2_Y}{6}(v^2_1-v^2_2)
       +\frac16 g^2_X (\frac16 v^2_1+\frac23 v^2_2-\frac56  u^2)
   , \nonumber
\end{eqnarray}
so the mass eigenvalues are given by,
\begin{eqnarray}
m^2_{\tilde t_\pm}&=&\frac12(m^2_{\tilde t_L}+m^2_{\tilde t_R})+y^2_tv^2_2
       \pm\sqrt{\frac14(m^2_{\tilde t_L}-m^2_{\tilde t_R})^2+y^2_tv^2_2
      (a_t-\lambda u\cot\beta)^2}, \nonumber \\
m^2_{\tilde b_\pm}&=&\frac12(m^2_{\tilde b_L}+m^2_{\tilde b_R})+y^2_bv^2_1
       \pm\sqrt{\frac14(m^2_{\tilde b_L}-m^2_{\tilde b_R})^2+y^2_bv^2_1
      (a_b-\lambda u\tan\beta)^2}, \nonumber\\
m^2_{\tilde \tau_\pm}&=&\frac12(m^2_{\tilde \tau_L}
        +m^2_{\tilde \tau_R})+y^2_\tau v^2_1
       \pm\sqrt{\frac14(m^2_{\tilde \tau_L}-m^2_{\tilde \tau_R})^2
        +y^2_\tau v^2_1
      (a_\tau-\lambda u\tan\beta)^2}, \nonumber\\
m^2_{\tilde g_\pm}&=&\frac12(m^2_{\tilde g}+m^2_{\tilde{g^c}})+k^2u^2
       \pm\sqrt{\frac14(m^2_{\tilde g}-m^2_{\tilde{g^c}})^2+
      (a_kku-\lambda kv_1v_2)^2} \nonumber.
\end{eqnarray}
The smuon has the same mass matrix structure as the stau. 
However the muon Yukawa
coupling is very small and the 
off-diagonal element of the smuon mass matrix is negligible. The typical
sparticle spectra are given in Table 2.

\begin{table}
\begin{center}
 \begin{tabular}{|l||c|c|c||c|c|c|} \hline
  $m_0$& $ 0(\eta)$ & $ 200(\eta)$ & $400(\eta)$ &
       $0(\xi)$ & $200(\xi)$ & $400(\xi)$ \\ \hline
  $k$              & 0.64  & 0.64   & 0.62   & 0.66  & 0.66   & 0.62\\
  $\lambda$        & 0.32  & 0.32   & 0.32   & 0.32  & 0.32   & 0.32\\
  $M_{30}$         & 220   & 220    & 220    & 220   & 220    & 220 \\
  $M_{20}$         & 800   & 800    & 750    & 800   & 750    & 700 \\
  $M_{Y0}$         & 900   & 800    & 500    & 900   & 1050   & 700 \\  \hline
  $Z'$             & 599.5 & 641.2  & 724.0  & 605.5 & 664.3  & 721.4\\
  $\tilde t_+$     & 601.2 & 631.6  & 696.8  & 568.9 & 584.0  & 643.6\\
  $\tilde t_-$     & 218.3 & 271.6  & 396.0  & 143.8 & 254.5  & 350.1\\
  $\tilde b_+$     & 558.1 & 592.3  & 664.3  & 527.2 & 546.6  & 624.4\\
  $\tilde b_-$     & 264.7 & 315.9  & 443.1  & 432.5 & 493.7  & 585.1\\
  $\tilde \tau_+$  & 447.6 & 478.1  & 544.4  & 566.5 & 602.8  & 665.6\\
  $\tilde \tau_-$  & 409.9 & 440.3  & 528.9  & 355.0 & 455.5  & 505.6\\
  $\tilde\nu_\tau$ & 440.7 & 471.7  & 538.7  & 561.2 & 597.8  & 661.0\\
  $\tilde g_+$     & 1071.3& 1138.7 & 1256.1 & 939.8 & 1029.9 & 1084.7\\
  $\tilde g_-$     & 729.5 & 803.4  & 922.2  & 602.9 & 687.5  & 736.1\\
  $g$              & 892.2 & 954.9  & 1046.0 & 763.7 & 838.1  & 854.9\\
  $\chi^\pm_1$     & 255.3 & 257.4  & 244.1  & 246.7 & 236.7  & 224.0\\
  $\chi^\pm_2$     & 467.4 & 496.3  & 554.5  & 400.7 & 429.7  & 460.4\\
  $h^0$            & 125.7 & 125.7  & 125.7  & 115.5 & 116.5  & 116.4\\
  $h^\pm$          & 442.1 & 502.9  & 661.6  & 309.0 & 380.9  & 455.9\\ \hline
 \end{tabular}
 \caption{The mass spectra of the extra U(1) models with $\tan\beta=20$
 and $A_0=0$ (GeV).}
\end{center}
\end{table}


%
%
%
\end{document}